\documentclass[12pt]{article}
\usepackage{hyperref}
\usepackage[dvipdfmx]{graphicx}
\usepackage{amsmath}
\usepackage{amssymb}
\usepackage{braket}
\usepackage{authblk}
\usepackage{mathrsfs}
\usepackage{slashed}
\usepackage{color}
\usepackage{cite}
\usepackage{subcaption}
\makeatletter
 
 \@addtoreset{equation}{section}
 \makeatother
\newcommand{\be}{\begin{equation}}
\newcommand{\ee}{\end{equation}}
\newcommand{\bea}{\begin{eqnarray}}
\newcommand{\eea}{\end{eqnarray}}
\newcommand{\beann}{\begin{eqnarray*}}
\newcommand{\eeann}{\end{eqnarray*}}
\newcommand{\nn}{\nonumber}
\newcommand{\ba}{\begin{array}}
\newcommand{\ea}{\end{array}}

\newcommand{\del}{\partial}

\newcommand{\rhot}{{\tilde{\rho}}}

\newcommand{\bse}{{\boldsymbol{e}}}

\newcommand{\tXi}{\widetilde{\Xi}}
\newcommand{\htheta}{\hat{\theta}}

\newcommand{\redx}{x}
\newcommand{\tn}{\tilde{n}}

\DeclareMathOperator{\diag}{diag}

\DeclareMathOperator{\Tr}{Tr}

\def\XXint#1#2#3{{\setbox0=\hbox{$#1{#2#3}{\int}$}
     \vcenter{\hbox{$#2#3$}}\kern-.5\wd0}}
\def\Xint#1{\mathchoice
   {\XXint\displaystyle\textstyle{#1}}%
   {\XXint\textstyle\scriptstyle{#1}}%
   {\XXint\scriptstyle\scriptscriptstyle{#1}}%
   {\XXint\scriptscriptstyle\scriptscriptstyle{#1}}%
   \!\int}
\def\dashint{\Xint-}

\title{Generalized Kazakov-Migdal Models on Graphs\\via Artin-Ihara $L$-function and Random Partitions}
\author[1,2]{So Matsuura\thanks{s.matsu@keio.jp}}
\author[2]{Kazutoshi Ohta\thanks{kazutoshi.ohta@mi.meijigakuin.ac.jp}}
\affil[1]{\it \small Hiyoshi Departments of Physics, 
and Research and Education Center for Natural Sciences,
Keio University, Yokohama, Kanagawa 223-8521, Japan}
\affil[2]{\it \small Institute for Mathematical Informatics, Meiji Gakuin University, Yokohama, Kanagawa 244-8539, Japan}

\date{}

\begin{document}
\maketitle

\vspace*{-1cm}

\begin{center}
{\bf Abstract}
\end{center}

We introduce Kazakov-Migdal (KM)-type gauge theories on graphs via the Artin-Ihara $L$-function, providing a unified description of %these models.
the models proposed in prior works.
Using harmonic analysis on the group manifold, we reformulate the KM-type model on the cycle graph as a random partition model governed by the Schur measure. 
We exactly solve the KM-type model on the cycle graph in the fundamental representation as the random partition model in the large $N_c$ limit and demonstrate that the Gross-Witten-Wadia phase transition occurs precisely when the limiting shape of the Young diagram touches the boundary of the allowed representation space. 
We further clarify that this phase transition is intimately related to Bose-Einstein condensation, and the strong/weak coupling duality possesses a natural combinatorial interpretation as the exchange between the Young diagram and its complement, reflecting the functional equation of the Artin-Ihara $L$-function. We also establish a definitive relationship between the eigenvalue density of the unitary matrix and the Maya diagram density of the random partitions by deriving a droplet picture from the spectral curve.

\newpage

\section{Introduction}

Lattice gauge theory is a powerful tool to investigate the non-perturbative aspects of gauge theory, and has been widely used in various fields of physics, such as high energy physics, condensed matter physics, and quantum information.
In this approach, one discretizes the continuous space-time into a regular lattice, which can be viewed as a special case of a graph.

We have recently proposed gauge theories on the graph by generalizing the Kazakov-Migdal (KM) model \cite{kazakov1993induced} and have investigated their properties \cite{Matsuura:2022dzl,Matsuura:2022ner,Matsuura:2023ova,Matsuura-Ohta:duality,matsuura2024functional, matsuura2025fermions}. 
These models consist of unitary matrices assigned to the edges and scalar fields placed at the vertices which transform in a certain representation of the gauge group. We have shown that the partition function of these models reduces to the graph zeta function weighted by unitary matrices.
In particular, when the scalar fields are in the fundamental representation, the model is called the fundamental Kazakov-Migdal (FKM) model, and we have shown that it universally exhibits the same phenomenon as the Gross-Witten-Wadia (GWW) phase transition in the unitary matrix model \cite{Gross:1980he, Wadia:1980cp}.
Furthermore, we have shown that the FKM model on regular graphs exhibits a strong/weak coupling duality that stems from the functional equation of the graph zeta function.

So far, we have investigated the properties of these models from the perspective of unitary matrix models. However, using harmonic analysis on the group manifold, we can also reformulate the partition function as a sum over random partitions via the Plancherel theorem. 
The purpose of this paper is to clarify the relation between the unitary matrix model and the random partition model, and to re-interpret the properties of the model from the viewpoint of the random partitions. 
We will show that the third-order GWW phase transition of the FKM model on the cycle graph emerges when the limiting shape of the Young diagram touches the boundary of the allowed region, and that the duality between a Young diagram and its complement in a rectangle reflects the functional equation of the graph zeta function.

The FKM model on a cycle graph can be viewed as a specific instance of a broader class of unitary matrix models. 
Specifically, this class of unitary matrix models is equivalent to a random partition model whose partition function is governed by the Schur measure. 
For such models, previous studies \cite{Dutta:2016byx,Chattopadhyay:2017ckc} 
suggested an intriguing relationship between the eigenvalue density of the unitary matrix and the density of the Maya diagram associated with Young diagrams by using the observation that the system exhibits a free fermion droplet structure. 
Concurrently, random partition models endowed with the Schur measure have been investigated by Kimura and Zahabi, who showed that the quantum spectral curve is intimately connected to the Maya density \cite{Kimura:2020sud,Kimura:2021lrc}. 
Bridging these developments, we demonstrate that the droplet structure naturally emerges from the spectral curve. 
From this perspective, we elucidate the explicit relationship between the eigenvalue density and the density of the Maya diagram, establishing that the FKM model serves as a concrete example of this framework.

This paper is organized as follows. 
In Sec.~2, after introducing the necessary notations and definitions of the graph theory, we review previous constructions of gauge theories on the graph in the literature. 
We unify them using the concept of the graph bundle and show that their partition functions can be systematically expressed in terms of the Artin-Ihara $L$-function.
In Sec.~3, we express the partition function of the unitary matrix model on the cycle graph as a sum over random partitions via the Plancherel formula.
In Sec.~4, we solve the random partition model associated with the FKM model on the cycle graph exactly in the large $N_c$ limit and show that the GWW phase transition emerges when the limiting shape touches the boundary of the allowed region. We also discuss the relation to Bose-Einstein condensation.
In Sec.~5, we discuss the strong/weak coupling duality of the KM-type model and interpret it as a duality replacing a Young diagram with its complement.
In Sec.~6, we establish the connection between the eigenvalue density and the Maya diagram density through the droplet structure and spectral curve analysis.
In the final section, we summarize the results and discuss future directions. 
Appendix A provides the detail of the exact solutions of the saddle point equation for the density at large $N_c$.

\section{Gauge Theories on the Graph and Graph Bundles}

In this section, we show that the KM-type gauge theories on the graph including the models proposed in \cite{Matsuura:2022dzl,Matsuura:2022ner,Matsuura:2023ova,Matsuura-Ohta:duality} can be understood in a unified way by using the concept of the graph bundle.

%%%
\subsection{Notations and the graph zeta functions}

We begin by fixing the graph theoretical notation used in this paper.
The graph is an object that consists of vertices $V$ and edges $E$ connecting them. 
In this paper, we consider only a simple graph which has at most one edge connecting two vertices and no edge connecting the same vertex.
In this case, we can define the edges of the graph through a symmetric mapping 
\begin{align}
  A: V\times V \to \{0,1\}\,, 
  \quad 
  A(v_1,v_2) = A(v_2,v_1)\,,
  \quad 
  A(v,v) = 0\,.
\end{align}
The set of edges $E$ is then given by the set of pairs $\{v_1,v_2\}$ of vertices which satisfy $A(v_1,v_2)=A(v_2,v_1)=1$ and we denote the graph determined by $V$ and $E$ as $\Gamma=(V,E)$.
We write the number of vertices and edges as $n_V=|V|$ and $n_E=|E|$, respectively.
Note that the mapping $A$ is represented by a square symmetric matrix of size $n_V$ called the adjacency matrix of the graph $\Gamma$, 
\begin{align}
  (A)_{v_1v_2} = \begin{cases}
    1 & \text{if $v_1$ and $v_2$ are connected by an edge}, \\
    0 & \text{otherwise}.
  \end{cases}
  \label{eq:adjacency matrix A}
\end{align}
In addition, we define the degree of a vertex $v$ as 
\begin{align}
  \deg v = \sum_{v'\in V} A(v,v')\,, 
\end{align}
which is the number of edges connected to $v$, 
and define the degree matrix $D$ as
\begin{align}
  D \equiv \diag(\deg v_1, \deg v_2, \cdots, \deg v_{n_V})\,.
\end{align}

In the definition above, the edge does not have a direction since the element of $E$ is given by a set of two vertices $\{v,v'\}$.
Such a graph is called an undirected graph.
We can give a direction to the edge by fixing the order of the vertices in each element of $E$ as $e\equiv \langle v,v'\rangle$, 
where $v=s(e)$ is called the source and $v'=t(e)$ is called the target of the directed edge $e$.
When the edges have directions, the graph is called a directed graph (digraph).
In the following, we assume that the edges are always directed and thus the graph means a digraph unless otherwise stated.
Corresponding to a directed edge $e$, we can consider the opposite directed edge $e^{-1}=\langle v',v\rangle$.
It is useful to enhance the set of the edges $E$ to
$E_D=\{\bse_a|a=1,\cdots,2n_E\}
=\{e_1,\cdots, e_{n_E}, e_1^{-1}, \cdots e_{n_E}^{-1}\}$,
combining with the opposite directed edges.

A path of length $\ell$ on the graph $\Gamma$ is a sequence of $\ell$ edges, $P=\bse_{a_1} \bse_{a_2} \cdots \bse_{a_\ell}$ satisfying the conditions $t(\bse_{a_i})=s(\bse_{a_{i+1}})$ ($i=1,\cdots,\ell-1$). 
We denote the length of the path $P$ by $\ell(P)$ and the starting and ending vertices of $P$ by $s(P)$ and $t(P)$, respectively.
The inverse of the path $P$ is defined as $P^{-1}=\bse_{a_\ell}^{-1}\cdots \bse_{a_1}^{-1}$. 
A path $C=\bse_{a_1} \cdots \bse_{a_\ell}$ that satisfies $s(\bse_{a_1})=t(\bse_{a_\ell})$ is called a cycle.
The product of two paths $P_1=\bse_{a_1} \cdots \bse_{a_{\ell_1}}$ and $P_2=\bse_{a'_1} \cdots \bse_{a'_{\ell_2}}$ can be defined when $t(\bse_{a_\ell})=s(\bse_{a'_1})$ as $P_1P_2 =\bse_{a_1} \cdots \bse_{a_{\ell_1}}\bse_{a'_1} \cdots \bse_{a'_{\ell_2}}$. 
In particular, a power of a cycle $C^n$, which is constructed through this product rule, is also a cycle.

A part of a path satisfying $\bse_{a_{j+1}}=\bse_{a_{j}}^{-1}$ is called a backtracking. 
If a cycle of length $\ell$ satisfies $\bse_{a_\ell}=\bse_{a_1}^{-1}$, this part is called a tail of the cycle $C$. 
For a cycle $C$, the equivalence class $[C]$ is defined by the set of cyclic rotations of $C$ as 
\[[C] = \{\bse_{a_1} \bse_{a_2} \cdots \bse_{a_\ell}, \,
\bse_{a_2} \cdots \bse_{a_\ell}\bse_{a_1}, \, \ldots \,,
\bse_{a_\ell}\bse_{a_1}\cdots\bse_{a_{\ell-1}}\}\,. \]
For the equivalence class $[C]$, backtracking and tail are indistinguishable, hence they are collectively called a bump.
The number of bumps on a cycle $C$ is denoted by $b(C)$. 
A cycle $C$ is called primitive when $C$ does not satisfy $C \ne B^r$ for any cycle $B$ and $r\ge 2$.
We denote the set of representatives of primitive cycles as $[{\mathcal P}]$.
A cycle $C$ is called reduced when it contains no bump. 
We denote the set of representatives of reduced cycles as $[{\mathcal P}_R] \subset [{\mathcal P}]$.

Among the graph zeta functions \cite{Ihara:original,MR607504,sunada1986functions,bartholdi2000counting}, 
the Bartholdi zeta function of a graph $\Gamma$ is defined by a product over the primitive cycles on $\Gamma$:
\begin{equation}
  \zeta_\Gamma(q,u)\equiv
  \prod_{[C]\in [{\cal P}]} \left({1-q^{\ell(C)} u^{b(C)} }\right)^{-1}
  =\exp\left(\sum_{[C]\in[{\cal P}]}\sum_{k=1}^\infty \frac{q^{\ell(C)k}u^{b(C)k}}{k}\right)\,. 
  \label{eq:Bartholdi}
\end{equation}
The Ihara zeta function is a special case of the Bartholdi zeta function with $u=0$. 
In this case, 
%Setting $u=0$ in \eqref{eq:Bartholdi}, 
such cycles with bumps do not contribute to the product in \eqref{eq:Bartholdi}, and the Ihara zeta function is given by a product over the primitive reduced cycles as
\begin{equation}
  \zeta_\Gamma(q)\equiv \zeta_\Gamma(q,0) =
  \prod_{[C]\in [{\cal P}_R]} \left(1-q^{\ell(C)}\right)^{-1}\,. 
  \label{eq:Ihara}
\end{equation}
One of the important properties of the Bartholdi zeta function is that it is expressed as the reciprocal of a polynomial using the adjacency matrix \eqref{eq:adjacency matrix A} of the graph as
\begin{align}
  \zeta_\Gamma(q,u) 
  =\left(1-q^2(1-u)\right)^{n_E-n_V} \det\left(I_{n_V}-q A + q^2(1-u)(D-(1-u)I_{n_V})\right)^{-1}\,. 
  \label{eq:Bartholdi zeta}
\end{align}
The essential reason why this equivalence holds is that both of $\ell(C)$ and $b(C)$ have the additivity such that
$\ell(C_1C_2)=\ell(C_1)+\ell(C_2)$,
$b(C_1C_2)=b(C_1)+b(C_2)$, 
which gives
\begin{align}
  q^{\ell(C_1C_2)} u^{b(C_1C_2)} = \left(q^{\ell(C_1)} u^{b(C_1)}\right) \left(q^{\ell(C_2)} u^{b(C_2)}\right)\,.
  \label{eq:additivity of length and bump}
\end{align}

%%%%
\subsection{Voltage assignment and the covering graph}

We consider a finite set $S$ and a bijection $\sigma_e: S\to S$ for each edge $e\in E$.
Since the bijection $\sigma_e$ can be regarded as a permutation of the elements of $S$, we can identify $\sigma_e$ as in an element of $\mathfrak{S}_{|S|}$. 
This is nothing but the assignment of a voltage $\sigma_e$ to the edge $e$, 
\begin{align}
  \alpha: E \ni e \mapsto \sigma_e \in \mathfrak{S}_{|S|}\,,
\end{align}
and is called the voltage assignment on the graph $\Gamma$.
We assume that the voltage assignment for the inverse edge $e^{-1}$ is the inverse of the voltage assignment for the original edge $e$; $\sigma_{e^{-1}}=\sigma_e^{-1}$.

Then, we consider the direct product of $V$ and $S$, $V_S \equiv V\times S$, and define a mapping $A_S^\alpha$ from $V_S\times V_S$ to $\{0,1\}$ by
\begin{align}
  A_S^\alpha((v_1,s_1),(v_2,s_2)) = 1 \quad \text{iff $A(v_1,v_2)=1$ and $\sigma_e(s_1)=s_2$},
  \label{eq:adjacency matrix AS}
\end{align}
for all $v_1,v_2\in V$ and $s_1,s_2\in S$.
This mapping determines a new graph $\Gamma_S^\alpha$ whose set of vertices is given by $V_S$ and the set of edges is determined by the mapping $A_S^\alpha$.
This graph is called the covering graph of $\Gamma$ with respect to the set $S$ and the voltage assignment $\alpha$.

In particular, 
if we choose the finite set $S$ to be a finite group $G$, 
the bijections $\sigma_e$ can be naturally given by the left multiplication of a certain element $g_e\in G$ as $\sigma_e(h)=g_e h$ for $h\in G$ 
by Cayley's theorem that any finite group $G$ is isomorphic to a subgroup of a symmetric group $\mathfrak{S}_{|G|}$. 
In this case, we can identify $\sigma_e$ with $g_e$ and  
the voltage assignment $\alpha$ is given by a mapping from $E$ to $G$ as
\begin{align}
  \alpha: E \ni e \mapsto g_e \in G\,. 
\end{align}
In this setting, the adjacency matrix $A_G^\alpha$ of the covering graph $\Gamma_G^\alpha$ is given by
\begin{align}
  (A_G^\alpha)_{(v,h)(v',h')} = \sum_{e\in E} \left( \delta_{v,s(e)}\delta_{v',t(e)} 
  \delta_{g_e h, h'} + \delta_{v,t(e)}\delta_{v',s(e)} \delta_{g_e^{-1} h, h'} \right)\,,
  \label{eq:adjacency matrix AG}
\end{align}
and we can define the Bartholdi zeta function of the covering graph $\Gamma_G^\alpha$ in the same way as \eqref{eq:Bartholdi zeta} by using the adjacency matrix \eqref{eq:adjacency matrix AG}.

%%%%
\subsection{The graph bundle and the Artin-Ihara $L$-function}

%The Bartholdi zeta function of the covering graph $\Gamma_G^\alpha$ 
%has an interesting property. 
%To see this, 

The idea of the covering graph can be further extended. 
%Instead of the group $G$ itself, 
Instead of the finite set $S$, 
we consider the representation space (Hilbert space) ${\cal H}_R$ of a certain representation $R$ of the group $G$ and define the direct product of the set of vertices $V$ and the representation space ${\cal H}_R$ as $V_{{\cal H}_R} \equiv V\times {\cal H}_R$.
Then, we can repeat the same discussion to construct the covering graph in this setting.

The obtained structure denoted by $\Gamma_{G,R}^\alpha$ is not a graph in the usual sense because ${\cal H}_R$ is a continuous space, but it is rather a generalization of the fiber (vector) bundle. 
In fact, the group $G$ and the voltage assignment $\alpha$ are regarded as the structure group and the connection on the bundle, respectively.
There is a natural projection map from the bundle $\Gamma_{G,R}^\alpha$ to the base graph $\Gamma$:
\be
\pi \, : \, \Gamma_{G,R}^\alpha \rightarrow \Gamma.
\ee
In this sense, we call $\Gamma_{G,R}^\alpha$ the graph bundle of $\Gamma$ (see e.g.~\cite{PISANSKI198312,MOHAR1988215,Kwak_Lee_1990}).

Under this setting, 
we can generalize the Bartholdi zeta function \eqref{eq:Bartholdi} on the graph $\Gamma$ to the so-called Artin-Ihara $L$-function 
%\eqref{eq:Bartholdi zeta},
\begin{align}
L_\Gamma(q,u;G,R,\alpha)
&\equiv\prod_{[C]\in [{\cal P}]}\det\left(I_{d_R}-\rho_R(g_C)q^{\ell(C)}u^{b(C)} \right)^{-1}\,, 
\label{eq:Artin-Ihara L-function definition}
\end{align}
where $g_C$ is the element of $G$ associated with the cycle $C$ defined by the product of $g_e$ for each edge $e$ in the cycle $C$ as
\begin{align}
  g_C \equiv \prod_{e\in C} g_e\,, 
\end{align}
and $\rho_R(g_C)$ is the representation matrix of $g_C$ in the representation $R$. 
Just as the Ihara zeta function is related to the Bartholdi zeta function, we can also define the Artin-Ihara $L$-function corresponding to the Ihara zeta function by setting $u=0$ in \eqref{eq:Artin-Ihara L-function definition}
\begin{align}
L_\Gamma(q;G,R,\alpha)
&\equiv\prod_{[C]\in [{\cal P}_R]}\det\left(I_{d_R}-\rho_R(g_C)q^{\ell(C)} \right)^{-1}\,. 
\end{align}

Since the representation matrix $\rho_R(g_C)$ has the multiplicative property $\rho_R(g_{C_1C_2})=\rho_R(g_{C_1})\rho_R(g_{C_2})$, similarly to $q^{\ell(C)}$ and $u^{b(C)}$ in \eqref{eq:additivity of length and bump},
the Artin-Ihara $L$-function can be expressed by the reciprocal of a polynomial 
\begin{align}
L_\Gamma(q,u;G,R,\alpha)
&=
\left(1-q^2(1-u)\right)^{(n_E-n_V)d_R} \nn \\
&\qquad \times
\det\left(I_{n_V d_R}
-q A_{G,R}^\alpha +q^2(1-u)(D-(1-u)I_{n_V d_R})\right)^{-1}
\,,
\label{eq:Artin-Ihara L-function}
\end{align}
where $A_{G,R}^\alpha$ is a weighted adjacency matrix of size $n_V d_R$ defined by 
\begin{align}
  (A_{G,R}^\alpha)_{(v,a)(v',b)} = \sum_{e\in E} \left( \delta_{v,s(e)}\delta_{v',t(e)} 
  \rho_R(g_e)_{ab} + \delta_{v,t(e)}\delta_{v',s(e)} \rho_R(g_e^\dag)_{ab} \right)\,.
  \quad 
  (a,b=1,\cdots,d_R)
  \label{eq:adjacency matrix A R}
\end{align}

It is important to note that the Artin-Ihara $L$-function is gauge invariant. 
To see this, consider a gauge transformation on the voltage assignment:
\begin{align}
  g_e \to h_{s(e)} g_e h_{t(e)}^{-1}\,,
  \label{eq:gauge transformation}
\end{align}
where $h_v\in G$ is assigned to each vertex $v\in V$. Since $g_C$ for any cycle $C$ is invariant under this transformation, the Artin-Ihara $L$-function (defined via $g_C$ in \eqref{eq:Artin-Ihara L-function definition}) is also invariant.

%%%%
\subsection{Decomposition of the Bartholdi zeta function of the covering graph}

The Artin-Ihara $L$-function \eqref{eq:Artin-Ihara L-function} and the Bartholdi zeta function of the covering graph $\Gamma_G^\alpha$ are closely related.

When the representation $R$ is reducible and decomposes into irreducible representations $R=\oplus_i R_i$, the Artin-Ihara $L$-function \eqref{eq:Artin-Ihara L-function} factorizes as
\be
L_\Gamma(q,u;G,R,\alpha)
=\prod_i L_\Gamma(q,u;G,R_i,\alpha)
\, ,
\ee
since the weighted adjacency matrix $A_{G,R}^\alpha$ can be decomposed as $A_{G,R}^\alpha=\oplus_i A_{G,R_i}^\alpha$ by taking an appropriate basis.
Therefore, 
if we choose $R$ to be the regular representation $R_{\rm reg}$, 
which includes all irreducible representations $R$ with multiplicity $d_R$, 
we obtain
\be
L_\Gamma(q,u;G,R_{\rm reg},\alpha)
=\prod_{R:\,\text{irr.rep.}} L_\Gamma(q,u;G,R,\alpha)^{d_R}
\,.
\label{eq:decomposition of regular L}
\ee
The representation space of the regular representation is the group ring of $G$ and the element of $\rho_{R_{\rm reg}}(g)$ is given by 
\begin{align}
  \rho_{R_{\rm reg}}(g)_{hh'} \equiv \delta_{g, h' h^{-1}}\,.
  \label{eq:regular rep element}
\end{align}
Comparing this with \eqref{eq:adjacency matrix AG} and \eqref{eq:adjacency matrix A R}, we find that the weighted adjacency matrix $A_{G,R_{\rm reg}}^\alpha$ coincides with the adjacency matrix $A_G^\alpha$ of the covering graph $\Gamma_G^\alpha$. 
Thus, the Artin-Ihara $L$-function with the regular representation $R_{\rm reg}$ equals the Bartholdi zeta function of the covering graph $\Gamma_G^\alpha$. This gives the decomposition of the zeta function $\zeta_{\Gamma_G^\alpha}(q,u)$ by the Artin-Ihara $L$-functions on the base graph $\Gamma$ \cite{sato2006weighted}:
\begin{align}
  \zeta_{\Gamma_G^\alpha}(q,u) 
  =  
  \prod_{R:\,\text{irr.rep.}} L_\Gamma(q,u;G,R,\alpha)^{d_R}\,.
  \label{eq:zeta decomposition}
\end{align}

%%%
\subsection{The Artin-Ihara $L$-function as the partition function of a KM-type gauge theory}

While the above discussion assumes $G$ to be a finite group, the same procedure can be extended to the case where $G$ is the compact Lie group $U(N_c)$.
In this case, we can construct a gauge theory on the graph bundle $\Gamma_{G,R}^\alpha$ where the representation $R$ and the voltage assignment $\alpha$ form the gauge field configuration. 

As a concrete example, we consider $N_f$ scalar fields in the representation $R$ of $U(N_c)$ as $\Phi_v^I\in {\cal H}_R$ on each vertex $v\in V$ 
where $I=1,\cdots,N_f$ is the flavor index. 
Mathematically speaking, the scalar fields $\Phi_v^I$ give a section of the graph bundle $\Gamma_{G,R}^\alpha$. 
We also regard the voltage assignment $\alpha$ on each edge $e$ as the gauge field (connection) $U_e\in U(N_c)$ on the graph bundle $\Gamma_{G,R}^\alpha$.
Then, we define the gauge invariant action on the graph $\Gamma$ by 
\be
S^\Gamma_R = \sum_{I=1}^{N_f}%\Tr\left\{
  \sum_{v\in V}m_v^2\Phi_v^{I\dag}\Phi_v^I
-q\sum_{e\in E}\left(\Phi_{s(e)}^{I\dag} \rho_R(U_e)\Phi^I_{t(e)}
+\Phi_{t(e)}^{I\dag} \rho_R(U_e^\dag)\Phi_{s(e)}^I
\right)\,,
%\right\}\, ,
\ee
where $q$ is a coupling constant and $m_v^2$ is the mass parameter at the vertex $v$.
This is a further generalization of the KM model \cite{kazakov1993induced} on the graph introduced in \cite{Matsuura:2022dzl,Matsuura:2022ner,Matsuura:2023ova} to an arbitrary representation.

Repeating the same calculation carried out in \cite{Matsuura:2022dzl,Matsuura:2022ner,Matsuura:2023ova}, we can show that the partition function of the model by tuning the mass parameters as 
\begin{align}
  m_v^2 = 1 - q^2(1-u)^2 + q^2(1-u)\deg v
  \label{eq:mass parameter}
\end{align}
is given by the unitary matrix integral of the Artin-Ihara $L$-function \eqref{eq:Artin-Ihara L-function}; 
\be
\begin{split}
Z^\Gamma_R &\equiv \int \prod_{v\in V}d\Phi_v\prod_{e\in E}dU_e\,
e^{- S}
=\int \prod_{e\in E}dU_e \,L_\Gamma(q,u;U(N_c),R,\alpha)^{N_f}\,,
\label{eq:Z by L}
\end{split}
\ee
with a Haar measure $dU_e$ of the group manifold $U(N_c)$.
This means that the Artin-Ihara $L$-function can be regarded as the partition function of the KM-type gauge theory on the graph bundle $\Gamma_{G,R}^\alpha$.

From this point of view,
the KM model on the arbitrary graph (gKM model) proposed in \cite{Matsuura:2022dzl,Matsuura:2022ner} is the model with the representation $R$ being the adjoint representation of $U(N_c)$, 
\be
  \rho_{\rm adj}(U_e) = U_e\otimes U_e^\dag\,. 
  \label{eq:rho adjoint}
\ee
This model is special because we assign a Hermitian matrix $\Phi_v$ (rather than a complex matrix) to each vertex $v$ as well as in the original KM model.
The partition function of the gKM model reduces to 
\begin{align}
  Z_{\rm gKM}^\Gamma=Z^\Gamma_{\rm adj} = \int \prod_{e\in E}dU_e \,|L_\Gamma(q,u;U(N_c),{\rm adj},\alpha)|^{\frac12}\,, 
  \label{eq:Z gKM}
\end{align}
where the factor $1/2$ in the exponent comes from the fact that the scalar fields $\Phi_v$ are Hermitian matrices and thus the number of degrees of freedom is half of that of complex matrices.

In the same way, the fundamental KM model (FKM model) considered in \cite{Matsuura:2023ova,Matsuura-Ohta:duality} is the model with the representation $R$ being the fundamental representation of $U(N_c)$,
we identify the element $U_e\in U(N_c)$ with the unitary matrix of size $N_c$ in the fundamental representation as
\be
  \rho_{\rm fund}(U_e) = U_e \, ,
  \label{eq:rho fundamental}
\ee
and the partition function becomes 
\begin{align}
   Z^\Gamma_{\rm FKM}=Z^{\Gamma}_{\rm fund} = \int \prod_{e\in E}dU_e \,|L_\Gamma(q,u;U(N_c),{\rm fund},\alpha)|^{N_f}\,.  
\end{align}

Before closing this section, we comment on the ``covering graph'' $\Gamma_{U(N_c),R}^\alpha$ with ``vertices'' $V_{U(N_c)}=V\times U(N_c)$ and ``edges'' defined by the mapping $A_{U(N_c)}^\alpha$ as in \eqref{eq:adjacency matrix AS}. 
Here the voltage assignment $\alpha$ maps edges to $U(N_c)$: $\alpha: E\ni e \mapsto U_e \in U(N_c)$.
Since the Lie group $U(N_c)$ has uncountably many elements, the covering graph $\Gamma_{U(N_c)}^\alpha$ is not a graph in the usual sense and it seems difficult to define the Bartholdi zeta function. 
However, since the Artin-Ihara $L$-function $L_\Gamma(q,u;U(N_c),R,\alpha)$ is defined for any representation $R$ of $U(N_c)$, we can {\it define} the Bartholdi zeta function of the covering graph $\Gamma_{U(N_c)}^\alpha$ by the right-hand side of the decomposition \eqref{eq:zeta decomposition} as
\begin{align}
  \zeta_{\Gamma_{U(N_c)}^\alpha}(q,u) 
  \equiv  
  \prod_{R\in Y_{N_c}} L_\Gamma(q,u;U(N_c),R,\alpha)^{d_R}\,,
\end{align}
where $Y_{N_c}$ is a set of all Young diagrams with maximal length $N_c$, which corresponds to the set of all irreducible representations $R$ of $U(N_c)$.

\section{Unitary Matrix Model and Random Partitions}
\label{sec:unitary to partitions}

In this section, we rewrite the partition function of the KM-type model on the cycle graph $C_L$ in terms of random partitions by using the harmonic analysis on the group manifold of $G=U(N_c)$.

%%%
\subsection{The Plancherel formula}

Let us consider a generic unitary matrix model with partition function
\be
Z =\int dU\, |f(U)|^2\, ,
\label{eq:Z general unitary model}
\ee
where $U$ is a unitary matrix of size $N_c$, $dU$ is the Haar measure over the group manifold of $G=U(N_c)$, and $f(U)$ is a class function.
In general, the class function $f(U)$ can be expanded in terms of the characters of the irreducible representations of the group $G$ as
\be
f(U)=\sum_{R\in Y_{N_c}} c_R\, \chi_R(U)\,,
\label{character expansion}
\ee
where $Y_{N_c}$ is a set of all Young diagrams with maximal length $N_c$ ($\ell(R)\leq N_c$) parametrized by the partition $R=\{\lambda_1\ge \lambda_2 \ge \cdots \lambda_{N_c}\}$
and the character of the representation $R$ of the group element $U$ is given by
\begin{align}
  \chi_R(U) = \Tr \rho_R(U) \equiv \Tr_R U\,.
\end{align}
Since the characters satisfy the orthogonality relations
\begin{align}
  \int dU \, \chi_{R}(U)\chi_{R'}(U^\dag)=\delta_{RR'}\,,
  \quad 
  \sum_{R\in Y_{N_c}} \chi_R(U)\chi_R(V^\dag) = \delta(U-V)\,,
\end{align}
the coefficient $c_R$ in \eqref{character expansion} is obtained from $f(U)$ by the integration 
\be
c_R=\int dU\, f(U)\chi_R(U^\dag)\, .
\ee
Because of the orthogonality of the characters, 
we see that 
the partition function \eqref{eq:Z general unitary model} is expressed as summation over the irreducible representations $R$ of the group $G$ as
\be
Z = \int dU\, |f(U)|^2 
  =
\sum_{R\in Y_{N_c}} |c_R|^2\, ,
\label{Plancherel model}
\ee
which is called the Plancherel formula.

%%%
\subsection{KM-type model on the cycle graph via random partitions}

Let us apply the Plancherel formula to the KM-type model on the cycle graph $C_L$ by setting $u=0$ for simplicity. 
Since the cycle graph $C_L$ has only two primitive reduced cycles, 
that is, a cycle of length $L$ that goes around the cycle in one direction and its inverse,
the Artin-Ihara $L$-function is given by 
\begin{align}
  L_{C_L}(q;U(N_c),R,\alpha)
  &= \left|\det\left(I_{\dim R} - q^L \rho_R(U)\right)\right|^{-2} \nn \\
  &= \exp\left(
    \sum_{k=1}^\infty \frac{q^{Lk}}{k}
    \left(\Tr_R U^k + \Tr_R U^{-k}\right)
  \right)\,,
  \label{eq:L cycle graph}
\end{align}
where $\dim R$ is the dimension of the representation $R$ of the gauge group $U(N_c)$ and $U$ is defined by the path-ordered product of the group elements assigned on the $L$ edges, 
$U\equiv U_{e_1}U_{e_2}\cdots U_{e_L}$.
In order for the expansion \eqref{eq:L cycle graph} to be well-defined, we need to assume $|q|<1$.
Since the Artin-Ihara $L$-function is gauge invariant, we can fix the gauge so that all the gauge fields $U_e$ except for one on a single edge, which is set to the identity matrix, and the partition function reduces to the unitary one-matrix model of a single unitary matrix $U$.
Therefore, 
the partition function of the model
\eqref{eq:Z by L} with $u=0$ 
is given by 
\begin{align}
Z^{C_L}(U(N_c),R) = \int dU\,\exp\left(
    N_f \sum_{k=1}^\infty \frac{q^{Lk}}{k}
    \left(\Tr_R U^k + \Tr_R U^{-k}\right)
  \right)\,,
  \label{eq:Z cycle graph}
\end{align}
in general. 
In this case, we restrict the coupling constant to $q < 1$ to ensure the partition function is well-defined.
This means that the function $f(U)$ corresponding to the KM-type model on the cycle graph $C_L$ is given by 
\begin{align}
f_R(U) \equiv \exp\left(
    N_f \sum_{k=1}^\infty \frac{q^{Lk}}{k}
      \Tr_R U^k
  \right)\,.
\end{align}

The character expansion of the function $f_R(U)$ is obtained for any representation $R$ of the gauge group $U(N_c)$ from \eqref{character expansion} in principle.  
Although we are interested in the case of the fundamental representation, namely, the FKM model on the cycle graph $C_L$ among them,
let us first consider the case of the adjoint representation, namely the gKM model on the cycle graph $C_L$:
\be
\begin{split}
f_{\rm adj}(U) 
%&=L_{C_L}(q;\rho_{\rm adj}(U))^{1/2}\\
&=\exp\left(
\frac12 \sum_{k=1}^\infty \frac{q^{L k}}{k}
|\Tr U^k|^2
\right)
=
\left(
\sum_{k=0}^\infty
q^{Lk}\sum_{\lambda\vdash k} \frac{1}{z_\lambda}
|\Upsilon_\lambda(U)|^2\right)^{\frac{1}{2}}
\, ,
\end{split}
\ee
where 
the trace without a subscript is assumed to be taken in the fundamental representation of $U(N_c)$,
$\lambda$ is a partition, $z_\lambda=\prod_{j\geq 1}j^{m_j}m_j!$ with $m_j$ the multiplicity of parts equal to $j$, and $\Upsilon_\lambda(U)$ is the string function defined by
\be
\Upsilon_\lambda(U) = \prod_{j=1}^\infty \left(\Tr U^j\right)^{m_j} \,,
\ee
which has the character expansion in terms of the irreducible representations $R$ of $U(N_c)$ as
\be
\Upsilon_\lambda(U) = \sum_{R} \chi_R({\cal C}_\lambda)\, \chi_R(U)\, .
\label{eq:string function character expansion}
\ee
Applying the Plancherel formula \eqref{Plancherel model} to the integral of the string function $\Upsilon_\lambda(U)$, we obtain
\begin{align}
  \int dU\, |\Upsilon_\lambda(U)|^2
  &= \sum_{R}\chi_R({\cal C}_\lambda)^2\, .
\end{align}
Combining it with the identity 
\begin{align}
  \sum_{\lambda\vdash m} \frac{1}{z_\lambda} \sum_{R}\chi_R({\cal C}_\lambda)^2
  = \mathfrak{p}_{N_c}(m)\,,
\end{align}
where $\mathfrak{p}_{N_c}(m)$ counts partitions of $m$ with the largest part at most $N_c$,
we can evaluate the partition function of the gKM model on the cycle graph $C_L$ as
\be
\begin{split}
Z_{\rm gKM}^{C_L}&=\int dU\, |f_{\rm adj}(U)|^2\\
&=\sum_{m=0}^\infty
q^{Lm}\sum_{\lambda\vdash m} \frac{1}{z_\lambda}
\int dU\, |\Upsilon_\lambda(U)|^2\\
&=\sum_{m=0}^\infty
\mathfrak{p}_{N_c}(m)q^{Lm}
=\prod_{i=1}^{N_c}\frac{1}{1-q^{Li}}\, ,
\label{partition function of adjoint}
\end{split}
\ee
which reproduces the result obtained from a direct calculation of the unitary matrix integral in \cite{Matsuura:2022dzl}.

We now consider the FKM model on the cycle graph $C_L$ with $N_f > N_c$ flavors. In this case, the GWW phase transition occurs in the large $N_c$ limit (see the next section). Since the partition function \eqref{eq:Z cycle graph} becomes 
\begin{align}
Z_{\rm FKM}^{C_L} = \int dU\,\exp\left(
    N_f \sum_{k=1}^\infty \frac{q^{Lk}}{k}
    \left(\Tr U^k + \Tr U^{-k}\right)
  \right)\,,  
  \label{eq:Z FKM on cycle graph}
\end{align}
the function $f(U)$ is given by 
\be
\begin{split}
f_{\rm fund}(U)
&=\exp\left(
\sum_{k=1}^\infty \frac{p_k(x)}{k}
\Tr U^k
\right)
=\sum_{\lambda}
\frac{p_\lambda(x)}{z_\lambda}
\Upsilon_{\lambda}(U)
\, ,
\end{split}
\ee
where the sum runs over all partitions $\lambda=\{\lambda_1\ge \lambda_2\ge \ldots\}$, 
$p_k(x)\equiv \sum_i x_i^k$ denotes the power sum 
$p_\lambda(x)\equiv p_{\lambda_1}(x)p_{\lambda_2}(x)\cdots$, 
and $x$ is specified by
\be
x=(\overbrace{q^L, \cdots ,q^L}^{N_f},0,\cdots)\, .
\label{specialization of x}
\ee
Using again the character expansion \eqref{eq:string function character expansion}, 
we can evaluate the coefficient $c_R$ for each irreducible representation $R$ as 
\be
\begin{split}
c_R&=
\sum_{\lambda} \frac{p_\lambda(x)}{z_\lambda}
\chi_R({\cal C}_\lambda)
%&=\frac{s_R(x)}{d_R}
=s_R(x)
\, ,
\label{eq:cR as Schur polynomial}
\end{split}
\ee
where $s_R(x)$ is the Schur polynomial of the variables $x$ associated with the representation $R$. 
Thus, substituting \eqref{eq:cR as Schur polynomial} into the Plancherel formula \eqref{Plancherel model}, we obtain the partition function
\be
\begin{split}
Z_{\rm FKM}^{C_L} &= \int dU \, |f_{\rm fund}(U)|^2
=\sum_{R\in Y_{N_c}} |s_R(x)|^2
\,,
\label{partition function of Schur functions}
\end{split}
\ee
which admits an interpretation as a random partition model with the Schur measure $\mathbb{P}^{\rm Schur}_R(x)\equiv |s_R(x)|^2$.
Therefore, the FKM model is a special case of random partition models with the Schur measure, which have been studied intensively in \cite{Kimura:2020sud,Kimura:2021lrc} (see \cite{marino2004houches} for a review).
In Sec.~\ref{sec:droplet and spectral curve}, we investigate this model from a general perspective. 

Recalling that the Schur function $s_R(x)$ is a sum of monomials over all semi-standard Young tableaux $T$ of the shape (Young diagram) $R$, where each power of $x_i$ counts the occurrences of the number $i$ in the box of $T$,
we obtain
\be
s_R(\overbrace{q^L, \cdots ,q^L}^{N_f},0,\cdots)
= (\dim_{N_f} R)\, q^{L|R|}\, ,
\ee
at the special values of the variables $x_i$ in the FKM model,
where $|R|$ is the number of boxes in the Young diagram of $R$, and
\be
\dim_{N_f} R \equiv 
\prod_{1\leq i < j \leq N_f}
\frac{\lambda_i-\lambda_j+j-i}{j-i}
\, ,
\ee
given by the Weyl character formula. 
We note that, although $\lambda_i$ are defined for $i=1,\cdots,N_f$ in this expression, we assume $\lambda_i=0$ for $N_c<i\leq N_f$ 
since $R$ belongs to a representation of the unitary group $U(N_c)$. 
Thus, we obtain another expression of the partition function of the FKM model on the cycle graph $C_L$ 
\be
Z_{\rm FKM}^{C_L}=\sum_{R\in Y_{N_c}} (\dim_{N_f} R)^2\, q^{2L|R|}\,.
\label{dual partition function}
\ee

We close this section by making a small comment on the case of $N_f=N_c$. 
In this case, $\dim_{N_f} R$ reduces to the dimension of the irreducible representation $R$ of $U(N_c)$ itself and the partition function becomes
\be
Z_{C_L}=\sum_{R\in Y_{N_c}} (\dim R)^2\, q^{2L|R|}\,, 
\label{Migdal partition function}
\ee
which coincides with the partition function of the two-dimensional BF theory on the sphere $S^2$, also known as the topological Yang-Mills theory \cite{Migdal:1975zg,Rusakov:1990rs} (see also \cite{Witten:1991we}).

%%%%%%%%%%%%%%%%%%%%%%%%%%%%%%%%%%%%%%%%%%%%%%%%%%%%%%%%%%%%
%%%%%%%%%%%%%%%%%%%%%%%%%%%%%%%%%%%%%%%%%%%%%%%%%%%%%%%%%%%%
\section{Exact Solution in the Large $N_c$ Limit}
%%%%%%%%%%%%%%%%%%%%%%%%%%%%%%%%%%%%%%%%%%%%%%%%%%%%%%%%%%%%
%%%%%%%%%%%%%%%%%%%%%%%%%%%%%%%%%%%%%%%%%%%%%%%%%%%%%%%%%%%%

In this section, we investigate the large $N_c$ behavior of the FKM model on the cycle graph $C_L$ from the view point of the random partition expression \eqref{dual partition function}.
We again consider only the case of $u=0$ for simplicity. 

\subsection{GWW phase transition on the cycle graph}

As a preliminary step, let us briefly review the properties of the model obtained in \cite{Matsuura:2023ova}, where it is investigated as a unitary matrix integral.

From the partition function \eqref{eq:Z FKM on cycle graph}, we see that the effective action of the FKM model on the cycle graph $C_L$ is given by that of a one unitary matrix model 
\begin{equation}
  S^{\rm eff}(q;U)=-\gamma N_c \sum_{k=1}^\infty \frac{q^{Lk}}{k}\Tr\left( U^k +  U^{-k}\right)\, ,
  \label{FKM on cycle graph}
\end{equation}
where $\gamma\equiv N_f/N_c$.
In order for the effective action to be well-defined, we need to assume $q<1$.
Although the effective action \eqref{FKM on cycle graph} includes higher order terms of the unitary matrix $U$ and its inverse $U^{-1}$, 
it reduces to the usual Wilsonian action 
\begin{equation}
  S^{\rm eff}_{C_L}(q;U)=-\gamma q^L N_c \Tr\left( U +  U^{-1}\right)\, ,
\label{GWW model}
\end{equation}
in the limit of $q\to 0$ with $\gamma q^L$ fixed, which is called the GWW model, that is, one plaquette reduction of two-dimensional Yang-Mills theory on the square lattice.

Despite the inclusion of infinitely many higher-order terms, the model (\ref{FKM on cycle graph}) can be solved exactly at large $N_c$ using the standard method for unitary matrix models.
If we diagonalize the unitary matrix $U$ as $U\to \diag(e^{i\theta_1},e^{i\theta_2},\cdots,e^{i\theta_{N_c}})$, the partition function of the FKM model on the cycle graph $C_L$ can be expressed in terms of an integral over the eigenvalues $\theta_i$ as
\begin{align}
  Z_{C_n} 
  &= {\cal N} \int_{-\pi}^{\pi} \prod_{i=1}^{N_c} d\theta_i\,
  e^{\sum_{j\ne k}\log\left|\sin \frac{\theta_j-\theta_k}{2}\right|
   -\gamma N_c\sum_{i} \log\left(1-2q^L \cos\theta_i + q^{2L} \right)
  }\,,
  \label{eq:ZCn}
\end{align}
with a normalization constant ${\cal N}$.

In the limit of $N_c\to\infty$ with fixing $\gamma = N_f/N_c$,
the density of the eigenvalues, 
\begin{equation}
  \rho(\theta)\equiv \frac{1}{N_c}\sum_{i=1}^{N_c}\delta(\theta-\theta_i)\,,
\end{equation}
becomes a continuous function as a solution of the saddle point equation.
In \cite{Matsuura:2023ova}, we showed that 
there is a critical value of the coupling $q$,  
%determined by the parameter $\gamma$ and is given by
\begin{equation}
  (q^*)^L = \frac{1}{2\gamma-1}\,,
  \label{eq:q critical}
\end{equation}
and  
the support of the density function $\rho(\theta)$ is a single connected interval $-\theta_0 \le \theta \le \theta_0$ ($0<\theta_0<\pi$) when $q>q^*$, 
whereas it covers the entire interval $-\pi\le \theta \le \pi$ when $q<q^*$. 
Since we assume $q<1$, we need to assume $\gamma>1$, that is, $N_f>N_c$ in order for the phase transition to occur.
The separation or merger of the support of the density function at a critical point is known as the GWW phase transition \cite{Gross:1980he,Wadia:1980cp}.
We can solve the saddle point equation satisfied by the density $\rho(\theta)$ as 
\begin{equation}
  \rho(\theta) = \begin{cases}\displaystyle
    \frac{1}{2\pi}\left(1+2\gamma\frac{q^L \cos\theta-q^{2L}}{1-2 q^L \cos\theta+q^{2L}}\right)\,, & \quad (0 < q \le q^*) \vspace{4mm}\\
    \displaystyle
    \frac{2(\gamma-1) q^L }{\pi}\frac{\cos\frac{\theta}{2}}{1-2 q^L \cos\theta+q^{2L}}\sqrt{\sin^2\frac{\theta_0}{2}-\sin^2\frac{\theta}{2}}\,, & \quad (q^*<q<1)
  \end{cases}
  \label{eq:sln}
\end{equation}
where $\theta_0$ is determined by 
\begin{equation}
  \sin^2\frac{\theta_0}{2} = \frac{(1-q^L)^2}{4q^L}\frac{2\gamma-1}{(\gamma-1)^2}\,, 
\end{equation}
and $\rho(\theta)=0$ outside of the support $-\theta_0\le \theta \le \theta_0$.

The free energy of the system is defined by
\begin{align}
  F_{C_L} &\equiv -\lim_{N_c\to\infty} \frac{1}{N_c^2}\log Z_{\rm FKM}^{C_L}\,.
  \label{eq:free energy}
\end{align} 
In order to evaluate the order of the phase transition, we define the ``internal energy'' 
\begin{equation}
  E_{C_L} \equiv \gamma \frac{\partial F_{C_L}}{\partial \gamma}\,,
  \label{eq:energy}
\end{equation}
the ``specific heat'' 
\begin{equation}
  C_{C_L} \equiv -\gamma^2 \frac{\partial^2 F_{C_L}}{\partial \gamma^2}\,, 
  \label{eq:specific heat}
\end{equation}
and the ``derivative of the specific heat''
\begin{equation}
  dC_{C_L} \equiv \gamma^3 \frac{\partial^3 F_{C_L}}{\partial \gamma^3}\,, 
  \label{eq:dC}
\end{equation}
where we treat $\gamma$ as inverse temperature.
From the eigenvalue density \eqref{eq:sln}, the free energy is evaluated as
\begin{equation}
F_{C_L} = \begin{cases}
\gamma^2 \log(1-q^{2L}) & (0< q \leq {q^*})\\
(2\gamma-1)\log(1-q^L)+\frac{1}{2}\log (2 \gamma q^L) +f(\gamma)
& ({q^*}< q < 1)
\end{cases}\, ,
\label{eq:free energy Cn}
\end{equation}
where
\begin{equation}
  f(\gamma) = 
  (\gamma-1)^2\log\left(1 - \frac{1}{\gamma}\right) 
  -2\left(\gamma - \frac{1}{2}\right)^2\log\left(1 - \frac{1}{2\gamma}\right) 
\end{equation}
is determined by the condition that the free energy is continuously connected at the critical coupling $q={q^*}$.
Substituting the expression \eqref{eq:free energy Cn} into the definitions \eqref{eq:energy}, \eqref{eq:specific heat}
and \eqref{eq:dC}, 
we obtain the internal energy
\begin{equation}
  E_{C_L} = \begin{cases}
    \displaystyle 
    2\gamma^2 \log(1-q^{2L}) &  (0< q \leq {q^*})\\
    \displaystyle 
    2\gamma \log(1-q^L) + e(\gamma)
    & ({q^*}< q < 1)
  \end{cases}\,,
  \label{eq:energy Cn}
\end{equation}
where 
\begin{equation}
  e(\gamma) \equiv 
  2\gamma \left(\gamma-1\right)\log\left(1 - \frac{1}{\gamma}\right)
  - 2\gamma(2\gamma - 1)\log\left(1 - \frac{1}{2\gamma}\right) \,,
\end{equation}
the specific heat 
\begin{equation}
  C_{C_L} = \begin{cases}
    \displaystyle -2\gamma^2 \log(1-q^{2L}) &  (0 < q \leq {q^*})\\
    \displaystyle -2\gamma^2 \log \frac{4\gamma(\gamma-1)}{(2\gamma-1)^2} & ({q^*} < q < 1)
  \end{cases}\,,
  \label{eq:specific heat Cn}
\end{equation}
and the derivative of the specific heat
\begin{equation}
  dC_{C_L} = \begin{cases}
    0 & (0< q \leq {q^*})\\
    \displaystyle \frac{2\gamma^3}{2\gamma^3 - 3\gamma^2 + \gamma}
    & ({q^*}< q < 1)
  \end{cases}\,. 
  \label{eq:dC Cn}
\end{equation} 
It is easy to check that the internal energy \eqref{eq:energy Cn} 
and the specific heat \eqref{eq:specific heat Cn}
connect continuously at $q={q^*}$ 
while the derivative of the specific heat does not. 
This means that the GWW phase transition of the FKM model on $C_L$ is third-order as the same as the original GWW model. For more details, see \cite{Matsuura:2023ova}. 

As a remark, we observe that the free energy in the small-$q$ regime is exactly reproduced by taking the large-$N_c$ limit of \eqref{partition function of Schur functions} while keeping $q$ and $N_f$ fixed, with $\gamma q^L \ll 1$.
In this limit,
since the Young diagram of the representation $R$ becomes unrestricted and runs over all partitions $\lambda$, we can apply the Cauchy-Littlewood identity
\be
\sum_\lambda |s_\lambda(x)|^2
=\prod_{i,j}\frac{1}{1-x_i\bar{x}_j} \,,
\label{eq:Cauchy-Littlewood identity}
\ee
which yields
\be
Z_{\rm FKM}^{C_L} = \frac{1}{\left(1-q^{2L}\right)^{N_f^2}}
\label{Cauchy-Littlewood partition function}
\ee
after setting $x$ to \eqref{specialization of x}. 
The free energy in this limit becomes
\be
F=\gamma^2\log(1-q^{2L})\, ,
\ee
which agrees with the result (\ref{eq:free energy Cn})
for smaller $q$ from the saddle point analysis of the unitary matrix model side.

\subsection{Limit to the Poissonized Plancherel measure model}

Before analyzing the random partition model in detail, 
we first point out that the model can be approximated as the so-called Poisonized Plancherel model in the region $N_f \gg N_c \gg 1$ and $q\ll 1$ with fixing $N_f q^L = \gamma N_c q^L$ to be finite.

We embed the irreducible representation $R$ of $U(N_c)$ into the irreducible representation of $U(N_f)$ with $N_f>N_c$ as 
\begin{align}
\lambda \equiv \{\lambda_1\ge \lambda_2\ge \cdots \ge \lambda_{N_c} \ge \lambda_{N_c+1} = \cdots = \lambda_{N_f}=0\}\,, 
\label{eq:partition by infinity lambdas}
\end{align}
where we have fixed the irrelevant $U(1)$ part of the representation by setting
$\lambda_j=0$ for $j>N_c$.
Then, we can express the dimension $\dim_{N_f} R$ as
\be
%\begin{split}
\dim_{N_f}R = \prod_{1\leq i < j \leq N_f}
\frac{\lambda_i-\lambda_j-i+j}{-i+j}
=\frac{d_\lambda}{|\lambda|!}
\prod_{(i,j)\in R}(N_f+j-i)
\, ,
\label{eq:dim Nf R}
%\end{split}
\ee
where $|\lambda|=\sum_{i}\lambda_i$ is the number of boxes in the Young diagram and 
\begin{align}
  d_\lambda 
  = |\lambda|! \prod_{1\le i < j < \infty} \frac{\lambda_i-\lambda_j+j-i}{j-i}
\end{align}
is the dimension of the irreducible representation of the symmetric group $\mathfrak{S}_{|\lambda|}$ associated with the partition $\lambda$ \cite{Nekrasov:2003rj}.
Since we can approximate 
$\prod_{(i,j)\in R}(N_f+j-i) \simeq {N_f}^{|\lambda|}$ 
when $N_f \gg N_c \gg 1$, 
the partition function (\ref{dual partition function}) in this region can be estimated as 
\be
Z^{C_L}_{\rm FKM} \simeq 
\sum_{\lambda} \left(\frac{d_\lambda}{|\lambda|!}\right)^2 \left(N_f q^{L}\right)^{2|\lambda|}
= 
e^{T^2}\sum_{\lambda}
\frac{d_\lambda^2}{|\lambda|!}
\frac{T^{2|\lambda|}e^{-T^2}}{|\lambda|!}
\, ,
\label{eq:Z as PPM}
\ee
where we have defined 
\begin{align}
  T\equiv N_f q^L\,.
\end{align}
Since we take the limit $N_c\to\infty$, the Young diagram $R$ is unrestricted and runs over all partitions $\lambda$.
This partition function can be regarded as a random partition model of the Poisson probability distribution with the Plancherel measure
\be
\mathbb{P}^{\rm Plancherel}_\lambda
=\frac{d_\lambda^2}{|\lambda|!}
\, .
\ee
The total measure, 
\be
\mathbb{P}^{\rm PPM}_\lambda(T)
=\frac{d_\lambda^2}{|\lambda|!}
\frac{T^{2|\lambda|}e^{-T^2}}{|\lambda|!}
\, ,
\ee
is called the Poissonized Plancherel measure.

Using the combinatorial identity
\be
\sum_{\lambda \vdash k}d_\lambda^2=k!\,, 
\ee
we can take the summation over the partitions $\lambda$ in \eqref{eq:Z as PPM} as 
\be
Z^{C_L}_{\rm FKM} \simeq e^{T^2}\sum_\lambda
\frac{d_\lambda^2}{|\lambda|!}
\frac{T^{2|\lambda|}e^{-T^2}}{|\lambda|!}
=e^{T^2}
=e^{N_c^2 \gamma^2 q^{2L}}
\, .
\ee
The free energy in this limit 
\be
F = -\frac{1}{N_c^2}\log Z^{C_L}_{\rm FKM} = -\gamma^2 q^{2L}
\ee
agrees with the result (\ref{eq:free energy Cn}) for small $q^L \ll 1$.

If the number of the boxes in the Young diagram $|\lambda|$ is very large, the Poissonized Plancherel measure behaves as
\be
\mathbb{P}^{\rm PPM}_\lambda(T)
\sim \exp\left\{
  -|\lambda| \log |\lambda| + |\lambda|  +2 |\lambda| \log T
\right\}
\, ,
\ee
which means that the expectation value of $|\lambda|$ is given by $T^2$ for large $|\lambda|$.
The asymptotic behavior of the Poissonized Plancherel measure is well investigated and 
it is known that the length of the longest increasing subsequences (LLIS) behaves as \cite{VK77,LOGAN1977206} (see also \cite{walsh:tel-03969540} for review)
\be
\ell(\lambda) \sim 2\sqrt{|\lambda|}
\sim 2T \, .
\ee
Recalling that there is a restriction $\ell(\lambda)\leq N_c$ in the original model (\ref{dual partition function}),
there is also a bound for the LLIS.
Thus, we expect that the phase transition occurs at the point
\be
2T^* = N_c \qquad
\Rightarrow
\qquad
(q^*)^L = \frac{1}{2\gamma}\, ,
\ee
for the large $\gamma$.
This agrees with the result (\ref{eq:q critical}) where $q^*$ is the critical value of the coupling $q$ at which the GWW phase transition occurs.

\subsection{Young diagrams, Maya diagrams and the FKM model by the random partitions}

\begin{figure}[htb]
\centering
\includegraphics[width=0.7\textwidth]{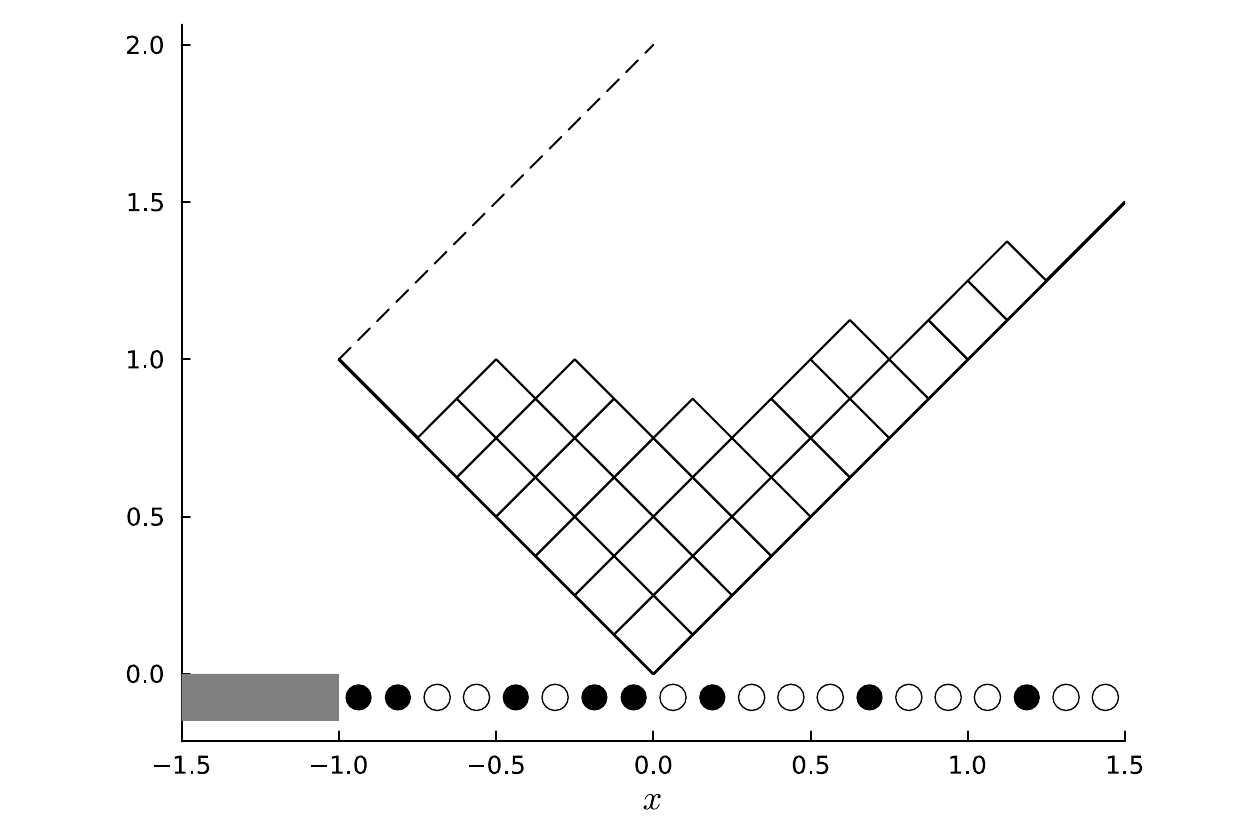}
\caption{The profile function and the Maya diagram corresponding to the partition $\lambda=(10,7,4,3,3,2,0,0)$ for $N_c=8$.
The black circles are points of the Maya diagram at $x_i=\epsilon n_i - \frac{\epsilon}{2}$, where $n_i$ is the non-colliding integers defined by $n_i=\lambda_i-i+1$. (The white circles are vacancies.) For the Young diagram for the representation of $U(N_c)$, $\ell(\lambda)\leq N_c$ is satisfied (a dashed line is its boundary), which means that the points should exist only at $x\geq -1$ ($n_{N_c}\geq - N_c+1$).
}
\label{fig:maya diagram}
\end{figure}
We next evaluate the partition function \eqref{dual partition function} in the large $N_c$ limit. 
The Young diagram is a graphical representation of a partition where $\lambda_i$ boxes are drawn in each row. The total number of boxes is $|\lambda|$.
%Although there are 
Among various ways to draw Young diagrams, 
we adopt the Russian convention where each row of the Young diagram is stacked at a 45-degree angle diagonally from the bottom right to the top left as drawn in Fig.~\ref{fig:maya diagram}. For later convenience, we rescale the size of each box in the Young diagram as $\sqrt{2}\epsilon$ with $\epsilon=1/N_c$.
The function that traces the shape of a Young diagram $\lambda$ is called the profile function, defined by 
\be
\begin{split}
  P_\lambda(x) &= |x|+\sum_{i=1}^{N_c}\left[
    |x-\epsilon(\lambda_i-i+1)|
    -|x-\epsilon(\lambda_i-i)|
    +|x+\epsilon i|
    -|x+\epsilon(i-1)|
  \right]\, ,
\end{split}
\ee
where $x$ is the horizontal coordinate of the profile function.
In the large $N_c$ limit ($\epsilon\to 0$), the profile function asymptotically approaches a smooth limiting shape
for sufficiently large $|\lambda|$.

The partition $\lambda$ is also expressed as the integers $n_i$ defined by  
\begin{align}
n_i \equiv \lambda_i-i+1\,,
\end{align}
which form a strictly decreasing sequence 
\be
n_1 > n_2 > \cdots > n_{N_c} \geq -N_c+1\,.
\label{integer sequence}
\ee
We see that the profile is downward sloping around the points
\be
x_i = \epsilon n_i - \frac{\epsilon}{2}
\, ,
\label{eq:maya diagram points}
\ee
and upward sloping otherwise. 
In Fig.~\ref{fig:maya diagram}, we draw the black circles at the points $x_i$ and the white circles at the other points. 
This expression based on the non-colliding integers $n_i$ is called the Maya diagram.

Let us now introduce the density of the points in the Maya diagram by
\be
\rhot(x) = \frac{1}{N_c}\sum_{i=1}^{N_c} \delta(x-x_i)\,.
\ee
From (\ref{integer sequence}) and the definition, the region of $x$ is bounded by $x\geq -1$ and $\rho(x)$ satisfies
\be
\int_{-1}^\infty \rhot(x) dx = 1\, .
\ee
Since points on the Maya diagram are lined up at interval $\epsilon$ even in the most dense state, the density should obey $\rhot(x)\leq 1$. 

Using $n_i$'s, $\dim_{N_f} R$ is expressed as
\be
\begin{split}
\dim_{N_f}R &= \prod_{1\leq i < j \leq N_f}
\frac{n_i-n_j}{j-i}\\
&= \prod_{1\leq i < j \leq N_c}
\frac{n_i-n_j}{j-i}
\prod_{i=1}^{N_c}
\prod_{j=N_c+1}^{N_f}
\frac{n_i+j}{j-i}
\\
&= \prod_{1\leq i < j \leq N_c}
\frac{n_i-n_j}{j-i}
\prod_{i=1}^{N_c}
\frac{(n_i+N_f)!(N_c-i)!}{(n_i+N_c)!(N_f-i)!}
\, .
\end{split}
\ee
Then, the partition function (\ref{dual partition function}) can be rewritten as
\be
Z_{C_L} = {\cal N}' \sum_{n_1 > n_2 > \cdots > n_{N_c} \geq -N_c+1}e^{-\tilde{S}[\vec{n};q]}\, ,
\ee
where ${\cal N}'$ is an irrelevant constant and 
\be
\begin{split}
\tilde{S}[\vec{n};q]
&= -\sum_{i\neq j}\log\left|n_i-n_j\right|
-2\sum_{i}\log \frac{(N_f+n_i)!}{(N_c+n_i)!}
-\log q^{2L} \sum_i n_i\\
&= -N_c^2\dashint dx dx' \, \log\left|x-x'\right|\rhot(x) \rhot(x')\\
&\qquad -2N_c \int dx\, \log\frac{(N_f+x/\epsilon)!}{(N_c+x/\epsilon)!}\, \rhot(x)
-N_c^2 \log q^{2L} \int dx \, x \rhot(x) \, .
\label{eq:action of n}
\end{split}
\ee

In the large $N_c$ limit, the density $\rhot(x)$ becomes a smooth function and obeys the following saddle point equation
\be
\dashint_{-1}^\infty \,
dx' \, \frac{\rhot(x')}{x'-x}
= \log\left(\frac{x+\gamma}{x+1}q^L\right)
\, ,
\label{eq:saddle point equation}
\ee
with the constraint $\rho(x)\leq 1$,
where we have used the Stirling formula
\be
\log N!\sim N\log N - N
\ee
at large $N$.

The saddle point equation (\ref{eq:saddle point equation}) can be solved exactly and the details are summarized in Appendix \ref{app:density}. 
The solution of the density $\rhot(x)$ is given by  
\eqref{eq:rho small q} in the region $0 < q\leq q^*$ 
and \eqref{eq:rho large q} in the region $q^* <q <1$ 
with $q^{*L} = \frac{1}{2\gamma-1}$ 
as expected from the analysis of the same model from the unitary matrix model side. 
%As we have seen in Appendix \ref{app:density}, 
As we see in Sec.~\ref{app:internal energy}, the internal energy in both phases completely coincides with that obtained from the unitary matrix model approach. 
Therefore, the specific heat and its derivative also agree with the result from the unitary matrix model. The phase transition at $q^{*L} = \frac{1}{2\gamma-1}$ is of the third-order, and the phenomenon occurring at this point is the GWW phase transition.

\begin{figure}[htb]
  \begin{minipage}[b]{0.32\textwidth}
    \centering
    \includegraphics[width=\textwidth]{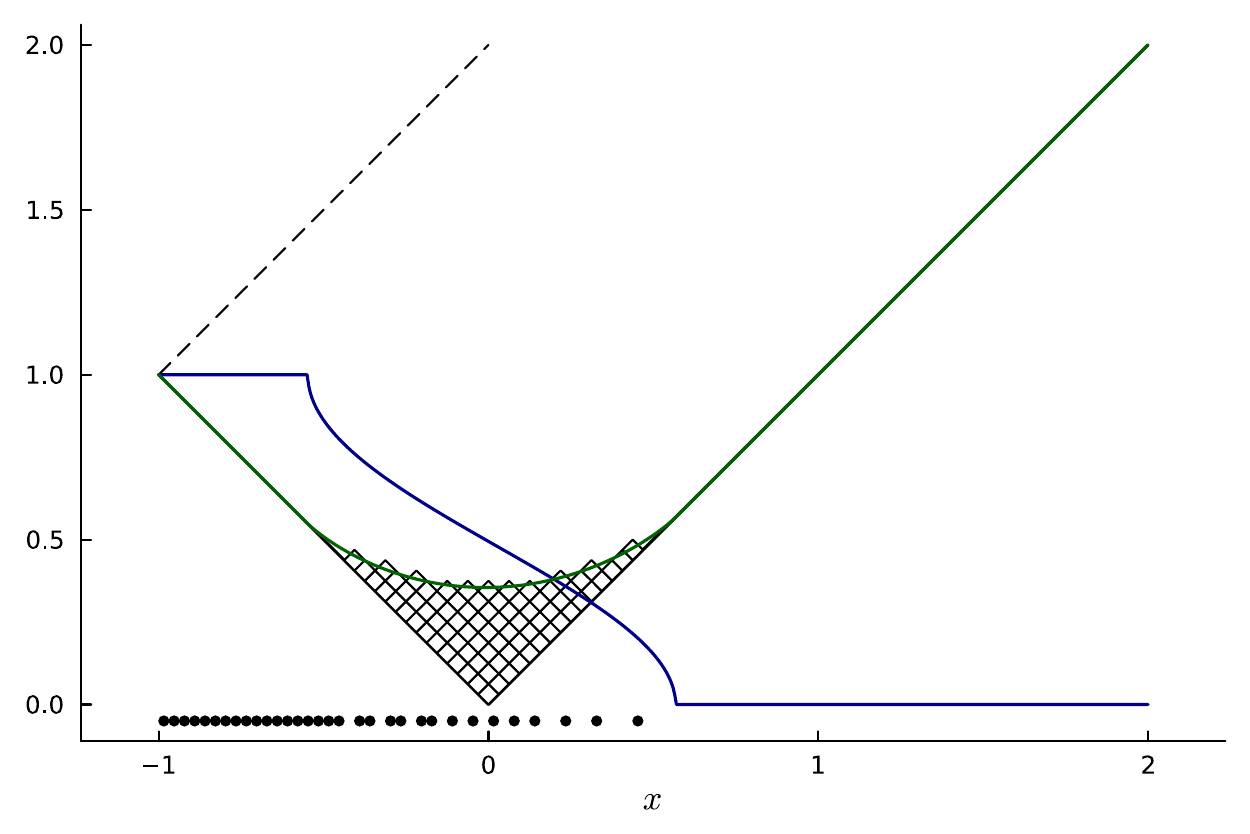}
    \subcaption{$q<q^*$}\label{q<qc}
  \end{minipage}
  \begin{minipage}[b]{0.32\textwidth}
    \centering
    \includegraphics[width=\textwidth]{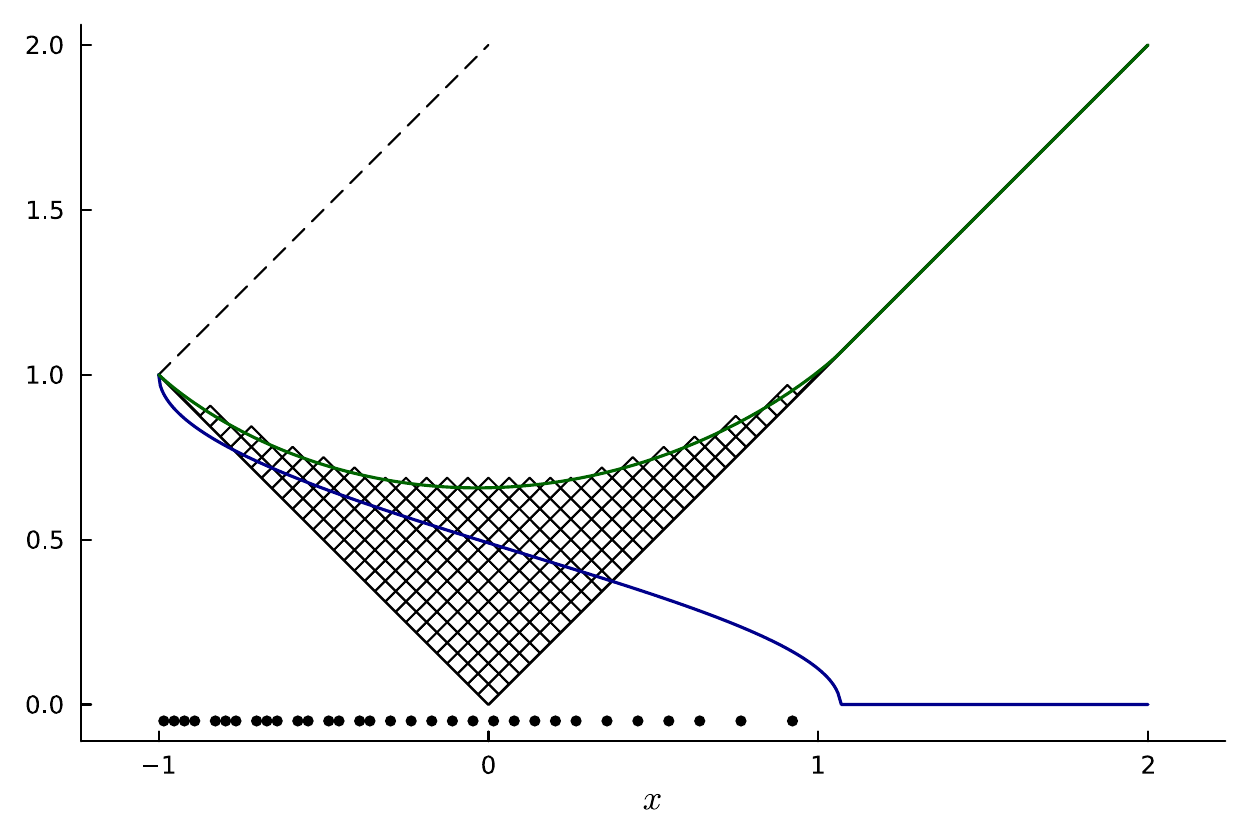}
    \subcaption{$q=q^*$}\label{q=qc}
  \end{minipage}
  \begin{minipage}[b]{0.32\textwidth}
    \centering
    \includegraphics[width=\textwidth]{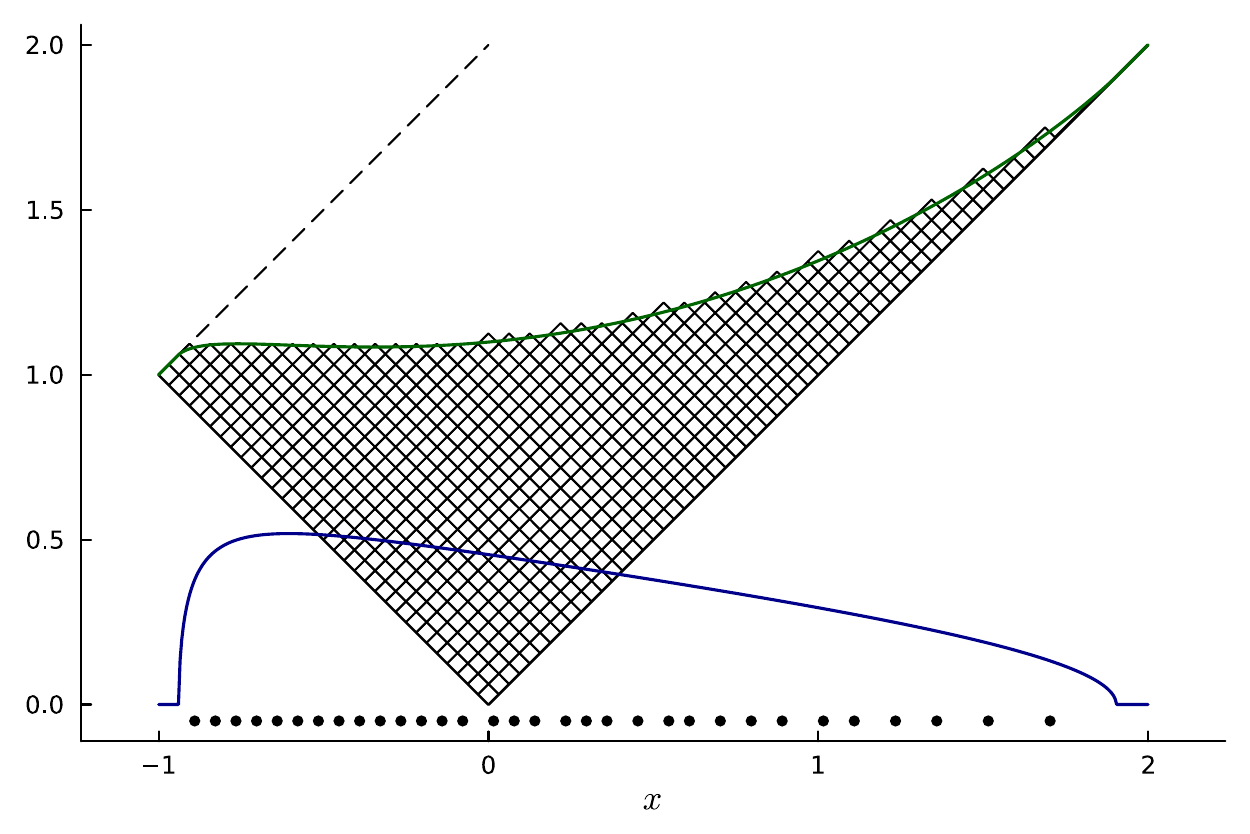}
    \subcaption{$q>q^*$}\label{q>qc}
  \end{minipage}
\caption{The blue lines indicate plots of the density $\rhot(x)$ of the Maya diagram points in various phases including near the phase transition point $q=q^*$, when $\gamma=16$. The green lines show the limiting shape of the profile of the Young tableau derived from the density. Using the large $N_c$ solution, we also plot a typical configuration of the Young diagram and Maya diagram for $N_c=32$. The black points represent the positions of the non-colliding integers $n_i$.}
\label{fig:density plots}
\end{figure}
%%%
In order to see the nature of the GWW phase transition from the partition point of view, we examine the behavior of the limiting shape of the profile function 
which is derived from the primitive function of the density as 
\be
P_\lambda(x) = x + 2\left(1- \chi(x)\right) \,,
\quad 
\chi(x) \equiv \int_{-1}^x dx' \, \rhot(x')\,.
\ee
We have plotted the density $\rhot(x)$ and the limiting shape of the profile function in Fig.~\ref{fig:density plots} for various values of $q$.
As shown in the figure, at the critical point $q=q^*$, the limiting shape of the profile function touches the boundary of the Young diagram due to the restriction that the number of rows cannot exceed $N_c$. 
When $q$ is much smaller than $q^*$, the profile function does not feel the presence of the boundary and the model behaves as if the Young diagram were unrestricted. 
This is why the model behaves like the Poissonized Plancherel model in this regime, as discussed in the previous subsection.
On the other hand, when $q$ is larger than $q^*$, the limiting shape of the profile function becomes a ``melting rectangle'', that is, a rectangle of height $N_c$ with a tail extending from its upper-right corner.

\subsection{Relation to Bose-Einstein condensation}
\label{sec:BEC}

For a given partition $R$, there is another expression of the partition which indicates the number of rows of a given length,
\be
R \to (1^{m_1}, 2^{m_2}, \ldots, j^{m_j}, \ldots)\,,
\ee
where $m_j$ is the number of rows of length $j$ in the Young diagram of $R$.
We here add the number of the rows of length zero $m_0$ defined by $m_0 = N_c - \ell(R)$, where $\ell(R)$ is the height of the Young diagram of $R$.
Using this expression, we can identify the partition $R$ with the state of a Bose system, where $m_j$ is the number of bosons occupying the equal spacing energy level $\varepsilon_j=j\varepsilon$ ($j=0,1,2,\ldots$).

From this point of view, the GWW phase transition can be regarded as BEC, where the number of the ground states $m_0$ becomes ${\cal O}(N_c)$ in the small $q$ phase, while it becomes zero in the large $q$ phase.
In fact, using the solutions \eqref{eq:rho small q} and \eqref{eq:rho large q} of the density $\rhot(x)$ in the both phases,
we can evaluate the number of the ground states $m_0$ as
\begin{align}
  m_0 = \begin{cases}
  N_c \left(1-\frac{2\gamma q^L}{1+q^L}\right) & \text{for $0 < q \le q^*$} \\
  0 & \text{for $q^* < q < 1$}
  \end{cases}
  \,.
\end{align}

The degeneracy of states proportional to $(\dim_{N_f} R)^2$ in the partition function (\ref{dual partition function}) is crucial for the occurrence of BEC in this model.
In fact, it is easy to show that a simplified model with a canonical partition function without $(\dim_{N_f} R)^2$
\begin{align}
  Z_{N_c} = \sum_{R\in Y_{N_c}}e^{-\beta |R| \varepsilon}
  = \sum_{\sum m_i=N_c}  e^{-\beta \sum_{j=0}^\infty m_j \varepsilon_j}
= \prod_{k=1}^{N_c}\frac{1}{1-e^{-\beta \varepsilon k}}
  \,,
  \label{1d boson partition function}
\end{align}
which describes one-dimensional $N_c$ bosons with equal spacing energy like a harmonic oscillator,
does not exhibit BEC at a finite temperature,
because of the IR divergence in the number of excited states. 
In this context, setting $q=e^{-\beta \varepsilon/L}$ makes the partition function \eqref{1d boson partition function} equivalent to that of the gKM model on the cycle graph \eqref{partition function of adjoint}.
This equivalence shows that the absence of the GWW phase transition in the gKM model on the cycle graph has the same origin as the absence of BEC in lower-dimensional bosonic systems.

The relation between BEC and the GWW phase transition (confinement/deconfinement phase transition) has been discussed in the QCD picture \cite{PhysRevD.102.096013,Hanada2021}. 
The FKM model on a general graph can be regarded as a generalization of the lattice gauge theory and the GWW phase transition is known to occur in a highly universal manner \cite{Matsuura:2023ova}. 
It is interesting to understand the relation and role of BEC both in the QCD and the random partition picture.

%%%%%%%%%%%%%%%%%%%%%%%%%%%%%%%%%%%%%%%%%%%%%%%%%%%%%%%%%%%%%%
\section{Duality of the KM-type Model}
\label{sec:duality}
%%%
\subsection{Duality of the KM-type model on the regular graph}

In \cite{Matsuura-Ohta:duality}, we showed that the FKM model with $u=0$ defined on a $(t+1)$-regular graph, namely a graph in which every vertex has degree $t+1$, possesses a strong/weak coupling duality in which $q$ and $\frac{1}{tq}$ are exchanged, due to the functional equation of the Artin-Ihara $L$-function for the fundamental representation. 
It has been proved in \cite{matsuura2024functional} that the functional equation of the Artin-Ihara $L$-function holds even for the fundamental representation with the bump parameter $u$.
Using the same calculation, we can extend this to the general representation $R$ as 
\begin{multline}
  L_\Gamma \left(\frac{1}{(1-u)(t+u)q},u;G,R,\alpha\right)  \\
  = 
  \Biggl(
    (-1)^{n_V} 
  \frac{(1-(1-u)^2q^2)^{n_E-\frac{n_V}{2}}\,(1-(t+u)q^2)^{\frac{n_V}{2}}}
  {(1-\frac{1}{(t+u)^2q^2})^{n_E-\frac{n_V}{2}}\,(1-\frac{1}{(1-u)q^2})^{\frac{n_V}{2}}}
  \Biggr)^{d_R}\,
  L_\Gamma(q,u;G,R,\alpha)\,.
  \label{eq:functional equation U Bartholdi}
\end{multline}
Combining this with the definition of the KM-type model \eqref{eq:Z by L}, we find that the KM-type model on the regular graph possesses the strong/weak coupling duality exchanging $q$ and $\frac{1}{(1-u)(t+u)q}$, as in the FKM model discussed in \cite{Matsuura-Ohta:duality}.

We can also extend the functional equation in $u$ proved in \cite{matsuura2024functional} to the general representation $R$ as 
\begin{equation}
 %\zeta_G(q,1-t-u;U) 
   L_\Gamma(q, 1-t-u; G,R,\alpha)  \nn \\
   = \left(
    \frac{\left(1-(1-u)^2q^2\right)^{n_E-n_V}}
    {\left(1-(t+u)^2q^2\right)^{n_E-n_V}}
  \right)^{d_R}
   L_\Gamma(q, u; G,R,\alpha)\,.
   \label{eq:functional equation U Bartholdi wrt u}
\end{equation}

%%%
\subsection{Duality of the FKM model on the cycle graph from random partitions}

Since the cycle graph $C_L$ is a $2$-regular graph, the duality of the KM-type model on the regular graph implies that the FKM model on the cycle graph $C_L$ exchanges $q$ and $q^{-1}$. 
To see this duality in the random partition model, let us define the dual parameter $\tilde{n}_i$ as
\begin{align}
  \tn_i \equiv -n_{N_c-i} - N_c - N_f - 1\,.
\end{align}
Substituting this into the effective action \eqref{eq:action of n}, we obtain
\begin{align}
  S[\vec{n};q] &= 
  -\sum_{i\neq j}\log\left|\tn_i-\tn_j\right|
  -2\sum_{i}\log \frac{(N_f+\tn_i)!}{(N_c+\tn_i)!}
  +\log q^{2L} \sum_i n_i + \text{const.} \nn \\
  &= S[\vec{\tn};q^{-1}] + \text{const.}\,.
\end{align}
This indicates the duality of the FKM model.

A key observation is that the dual parameter $\tilde{n}_i$ corresponds to the non-colliding integers of the complement of the Young diagram $R$ in a rectangle.
Therefore, the duality of the FKM model on the cycle graph can be understood as a duality that replaces a Young diagram with its complement in the $N_c$-height rectangle. 
Similar duality appears in the string-theoretic interpretation of two-dimensional Yang-Mills theory \cite{gross1993two}.
It would be worthwhile to explore the relationship between these two descriptions further.

%%%%%%%%%%%%%%%%%%%%%%%%%%%%%%%%%%%%%%%%%%%%%%%%%%%%%%%%%%%%%%
\section{Droplet Picture and Spectral Curve}
\label{sec:droplet and spectral curve}

In this section, we will discuss the relation between the eigenvalue density of the unitary matrix model and the density of the non-colliding integers $n_i$ in the random partition model in a general framework including the FKM model on the cycle graph $C_L$.
The relation between the two densities has been discussed in \cite{Dutta:2016byx, Chattopadhyay:2017ckc}
based on the droplet picture of the phase space of the underlying free fermion system. 
We will see that the boundary equation of the droplet can be derived from the spectral curve analyzed in \cite{Kimura:2020sud,Kimura:2021lrc},
which gives a proof of the relation between the two densities in a more general framework.

%%%
\subsection{Character expansion and Schur-Weyl duality}

We consider the unitary matrix model with the action
\begin{align}
  S(\vec{\beta};U) \equiv -N_c \sum_{n=1}^\infty \frac{\beta_n}{n} \left(\Tr U^n + \Tr U^{-n}\right)\,, 
  \label{eq:general model action}
\end{align}
where $\vec{\beta} = (\beta_1, \beta_2, \cdots)$ is a set of coupling constants.
In the case of the FKM model on the cycle graph $C_L$, the coupling constants are given by $\beta_n = \gamma q^{nL}$.
In what follows, we evaluate the partition function $Z\equiv \int [dU] e^{-S(\vec{\beta};U)}$ by following the approach of \cite{Dutta:2016byx, Chattopadhyay:2017ckc}.

Repeating the same calculation in Sec.~\ref{sec:unitary to partitions}, 
the partition function of the unitary matrix model can be expressed as a sum over the irreducible representations $R$ of $U(N_c)$ as
\begin{align}
  Z \equiv \sum_{R\in Y_{N_c}} \bigl|\Xi_R(\vec{\beta})\bigr|^2\,,
  \label{eq:Z}
\end{align}
where $\Xi_R(\vec{\beta})$ is defined by 
\begin{align}
  \Xi_R(\vec{\beta}) \equiv 
  \sum_{m_1,m_2,\cdots=0}^\infty \prod_{k=1}^\infty \frac{1}{m_k!}\left(\frac{N_c \beta_k}{k}\right)^{m_k} \chi_{R}({\cal C}_{\vec{m}})\,,
  \label{eq:X_R}
\end{align}
where
${\cal C}_{\vec{m}}$ is the conjugacy class of the symmetric group $\mathfrak{S}_{|R|}$ specified by the cycle structure $\vec{m}=(m_1,m_2,\cdots)$, 
$m_k$ is the number of cycles of length $k$,
and $|R| = \sum_{k=1}^\infty k m_k$ is the total number of boxes in the Young diagram corresponding to the representation $R$.
Note that we do not need to restrict the summation of $m_k$ to satisfy $|R| = \sum_{k=1}^\infty k m_k$ in the definition of $\Xi_R(\vec{\beta})$ since $\chi_R({\cal C}_{\vec{m}})=0$ for $|R| \ne \sum_{k=1}^\infty k m_k$.

The representation $R$ can be expressed by a Young diagram with at most $N_c$ rows. 
We write the length of the rows of the Young diagram $R$ as $\lambda_1 \ge \lambda_2 \ge \cdots \ge \lambda_{N_c} \ge 0$ 
and define 
\begin{align}
  h_i \equiv \lambda_i - i + N_c = n_i + N_c\,.
\end{align}
Then, the character $\chi_R({\cal C}_{\vec{m}})$ can be expressed in terms of the Frobenius formula as
\begin{align}
  \chi_R({\cal C}_{\vec{m}}) = \left[
    \left(
    \prod_{i<j}^{N_c}(a_i-a_j)
    \right)
    \prod_{k=1}^\infty \left( \sum_{i=1}^{N_c} a_i^{k} \right)^{m_k}
  \right]_{(h_1,\cdots,h_{N_c})}\,,
\end{align}
where the subscript $(h_1,\cdots,h_{N_c})$ means that we take the coefficient of $a_1^{h_1} \cdots a_{N_c}^{h_{N_c}}$ in the expansion of the expression in the square brackets, 
which is achieved by the contour integral in the complex $z_i$-plane as 
\begin{align}
  \chi_R({\cal C}_{\vec{m}}) = \prod_{i=1}^{N_c} \oint \frac{dz_i}{2\pi i z_i^{h_i+1}} 
  \left(
    \prod_{i<j}^{N_c}(z_i-z_j)
    \prod_{k=1}^\infty \left( \sum_{i=1}^{N_c} z_i^{n} \right)^{m_k}
  \right)\,,
  \label{eq:character by contour integral}
\end{align}
  where the contour of the integral must surround the origin $z_i=0$ for each $i$, which is safely achieved by 
\begin{align}
  z_i \equiv e^{i\theta_i}
  \quad 
  (\theta_i \in [-\pi, \pi])\,.
  \label{eq:z by theta}
\end{align}

Substituting the expression of the character \eqref{eq:character by contour integral} into the definition of $\Xi_R(\vec{\beta})$ \eqref{eq:X_R}, we can perform the summation over $m_k$ and obtain
\begin{align}
  \Xi_R(\vec{\beta}) &= 
  \oint \prod_{i=1}^{N_c}\frac{dz_i}{2\pi i z_i}
  \exp\left(
    N_c \sum_{i=1}^{N_c}\sum_{k=1}^\infty \frac{\beta_k}{k}z_i^k
    + \frac12 \sum_{i\ne j=1}^{N_c}\log(z_i-z_j) 
    - \sum_{i=1}^{N_c} h_i \log z_i
  \right) \nn \\
  &= 
   \int_{-\pi}^\pi \prod_{i=1}^{N_c}\frac{d\theta_i}{2\pi} 
  \exp\Bigg(
    N_c \sum_{i=1}^{N_c}\sum_{k=1}^\infty \frac{\beta_k}{k}e^{i k \theta_i}
    + \frac12 \sum_{i\ne j=1}^{N_c}\log\left|2\sin\frac{\theta_i-\theta_j}{2}\right|\nn\\
    &\qquad\qquad\qquad\qquad\qquad
    - i \sum_{i=1}^{N_c} \left(h_i - \frac{N_c-1}{2}\right) \theta_i
  \Bigg) \nn \\
  &\equiv \Xi(\vec{h};\vec{\beta})\,, 
  \label{eq:X_R by Fourier transform}
\end{align}
where we have renamed the function $\Xi_R(\vec{\beta})$ as $\Xi(\vec{h};\vec{\beta})$ to emphasize that it is a function of $\vec{h} = (h_1, \cdots, h_{N_c})$.

It is important to notice that $\Xi(\vec{h};\vec{\beta})$ is Fourier transformation of the function of 
$\vec{\theta} = (\theta_1, \cdots, \theta_{N_c})$; 
\begin{align}
  \tXi(\vec{\theta};\vec{\beta}) \equiv \exp\left(
    N_c \sum_{i=1}^{N_c}\sum_{k=1}^\infty \frac{\beta_k}{k}e^{i k \theta_i}
    + \frac12 \sum_{i\ne j=1}^{N_c}\log\left|2\sin\frac{\theta_i-\theta_j}{2}\right|
  \right)\,. 
\end{align}
This means that the partition function \eqref{eq:Z} can be expressed from the Plancherel theorem as
\begin{align}
  Z = \sum_{\vec{h}} \bigl|\Xi(\vec{h};\vec{\beta})\bigr|^2 = \int_{-\pi}^\pi \prod_{i=1}^{N_c} \frac{d\theta_i}{2\pi}\, \bigl|\tXi(\vec{\theta};\vec{\beta})\bigr|^2\,.
  \label{eq:Z by Fourier transform}
\end{align}
This is an explicit demonstration of the Schur-Weyl duality between the unitary group $U(N_c)$ and the symmetric group $S_{|R|}$ in the context of the unitary matrix model and the random partition model.
In fact, we can write down \eqref{eq:Z by Fourier transform} as 
\begin{align}
  Z &= \int_{-\pi}^\pi \prod_{i=1}^{N_c} \frac{d\theta_i}{2\pi}\, \bigl|\tXi(\vec{\theta};\vec{\beta})\bigr|^2 \nn \\
  &= \int_{-\pi}^\pi \prod_{i=1}^{N_c} \frac{d\theta_i}{2\pi}\, \exp\left(
    2N_c \sum_{i=1}^{N_c}\sum_{k=1}^\infty \frac{\beta_k}{k}\cos k \theta_i
    + \sum_{i\ne j=1}^{N_c}\log\left|2\sin\frac{\theta_i-\theta_j}{2}\right|
  \right)\,,
\end{align}
which reproduces the partition function of the unitary matrix model with the action $S(\vec{\beta};U)$.
This means that the parameter $z_i$ given by \eqref{eq:z by theta} can be regarded as the eigenvalues of the unitary matrix $U$ in the unitary matrix model \cite{Dutta:2016byx}.

%%%
\subsection{Schwinger-Dyson equation and droplet picture}

Let us now consider the Schwinger-Dyson equation with respect to \eqref{eq:X_R by Fourier transform}:
\begin{align}
  0 &= \int_{-\pi}^\pi \prod_{j=1}^{N_c}\frac{d\theta_j}{2\pi} 
  \frac{\del}{\del \theta_i}
  \exp\Bigg(
    N_c \sum_{j=1}^{N_c}\sum_{k=1}^\infty \frac{\beta_k}{k}e^{i k \theta_j}
    + \frac12 \sum_{j\ne k=1}^{N_c}\log\left|2\sin\frac{\theta_j-\theta_k}{2}\right|\nn\\
    &\qquad\qquad\qquad\qquad\qquad\qquad
    - i \sum_{j=1}^{N_c} \left(h_i - \frac{N_c-1}{2}\right)\theta_j
  \Bigg) \nn \\
  &= \int_{-\pi}^\pi \prod_{j=1}^{N_c}\frac{d\theta_j}{2\pi} 
  \left(
    iN_c\sum_{n=1}^\infty \beta_n e^{in\theta_i} 
    +\frac12 \sum_{j\ne i}\cot\frac{\theta_i-\theta_j}{2} -i \left(h_i - \frac{N_c-1}{2}\right)
  \right) \nn \\
  &\qquad\times \exp\left(
    iN_c \sum_{j=1}^{N_c}\sum_{k=1}^\infty \frac{\beta_k}{k}e^{i k \theta_j}
    + \frac12 \sum_{j\ne k=1}^{N_c}\log\left|2\sin\frac{\theta_j-\theta_k}{2}\right|
    - i \sum_{j=1}^{N_c} \left(h_j - \frac{N_c-1}{2}\right) \theta_j
  \right) \nn \\
  &\equiv N_c \left\langle 
    i\sum_{k=1}^\infty \beta_k e^{ik\theta_i} 
   +\frac{1}{2N_c}\sum_{j\ne i}\cot\frac{\theta_i-\theta_j}{2} - i\left(\frac{h_i}{N_c} - \frac12 + \frac{1}{2N_c}\right) \right\rangle\,,
  \label{eq:SD equation finite N}
\end{align}
which is satisfied even for finite $N_c$. 
Introducing the density of the eigenvalues $\theta_i$ as
\begin{align}
  \rho(\theta) \equiv \frac{1}{N_c}\sum_{i=1}^{N_c} \delta(\theta-\theta_i)\,,
\end{align}
we can rewrite the Schwinger-Dyson equation as
\begin{align}
  \left\langle 
    i \sum_{k=1}^\infty \beta_k e^{ik\theta_i} 
    + \frac12 \int_{C_i} d\theta \rho(\theta)
      \cot\frac{\theta_i-\theta}{2} 
      -i \left(\frac{h_i}{N_c} - \frac12 + \frac{1}{2N_c}\right)
      %- i\frac{\redn_i}{N_c} 
      \right\rangle = 0\,,
  \label{eq:SD eq pre}
\end{align}
where $C_i$ is the contour of the $\theta$ integral that avoids the singularity at $\theta=\theta_i$.
If we express the $\theta$ integral in the complex $z$-plane, the contour $C_i$ is obtained by taking $z=e^{i\theta}$ ($-\pi<\theta\le\pi$) and avoiding the point $z_i$ by moving it slightly inside or outside the unit circle $|z|=1$.
Therefore, the equation \eqref{eq:SD eq pre} can be rewritten as
\begin{align}
  0 &= \left\langle 
    i\sum_{k=1}^\infty \beta_k e^{ik\theta_i} 
    + \frac12 \int_{-\pi}^\pi d\theta \rho(\theta)
      \cot\frac{\theta_i-\theta\pm i\epsilon}{2}
    %-i \frac{\redn_i}{N_c} 
    -i \left(\frac{h_i}{N_c} - \frac12 + \frac{1}{2N_c}\right)
    \right\rangle  \nn \\
  &= \left\langle 
    i\sum_{k=1}^\infty \beta_k e^{ik\theta_i} 
    + \frac12 \dashint_{-\pi}^\pi d\theta \rho(\theta)
      \cot\frac{\theta_i-\theta}{2} 
    \pm \pi i \rho(\theta_i)
    -i \left(\frac{h_i}{N_c} - \frac12 + \frac{1}{2N_c}\right) \right\rangle\,,
\end{align}
where the ambiguity of the sign $\pm$ corresponds to the choice of the shift of $z_i$ inside or outside the unit circle%
\footnote{
  As we will mention soon, this is the different point of our argument from the one in \cite{Dutta:2016byx, Chattopadhyay:2017ckc}. 
}.

In the large $N_c$ limit, we define continuous variables from the variables $h_i$ and $\theta_i$ as 
\begin{align}
  t \equiv \frac{i}{N_c}\,, \quad x(t) \equiv \frac{h_i}{N_c}-1\,, \quad \theta(t) \equiv \theta_i\,.
\end{align}
Using these continuous variables, 
the Schwinger-Dyson equation becomes the semi-classical equation for the density $\rho(\theta)$ as%
\begin{align}
  i\sum_{k=1}^\infty \beta_k e^{ik\theta} 
  + \frac12 \dashint_{-\pi}^\pi d\theta' \rho(\theta')\cot\frac{\theta-\theta'}{2} 
  \pm \pi i \rho(\theta)
  - i \left(x + \frac12\right) = 0\,.
  \label{eq:large N SD}
\end{align}
The real part of this equation is 
\begin{align}
    \dashint_{-\pi}^\pi d\theta' \rho(\theta')\cot\frac{\theta-\theta'}{2} 
    = 2\sum_{k=1}^\infty \beta_k \sin k\theta \,,
  \label{eq:large N SD real}
\end{align}
which is precisely the saddle point equation that determines the eigenvalue density $\rho(\theta)$ as a function of $\vec{\beta}$ in the large $N_c$ limit. 
On the other hand, the imaginary part of the equation \eqref{eq:large N SD} is 
\begin{align}
    x + \frac12 = 
  \sum_{k=1}^\infty \beta_k \cos k\theta 
  \pm \pi \rho(\theta)\,.
  \label{eq:large N SD imag}
\end{align}
Substituting the solution $\rho(\theta)$ of \eqref{eq:large N SD real} into \eqref{eq:large N SD imag}, this equation allows us to determine the relation between $x$ and $\theta$ as
\begin{align}
  \left(x + \frac12 - \sum_{k=1}^\infty \beta_k \cos k\theta \right)^2 = \pi^2 \rho(\theta)^2\,, 
  \label{eq:boundary equation}
\end{align}
which is the equation of the boundary of the droplet in the phase space of the underlying free fermion system \cite{Dutta:2016byx, Chattopadhyay:2017ckc}. 
The area of the region in the $x$-$\theta$ plane surrounded by the curve \eqref{eq:boundary equation} is $2\pi$ because of the normalization condition $\int_{-\pi}^\pi d\theta \rho(\theta) = 1$.
We will see in the next subsection the origin of this area from the quantum mechanical point of view. 

We emphasize that we do not assume the droplet structure a priori. 
In the analyses of \cite{Dutta:2016byx,Chattopadhyay:2017ckc}, the term $\pm \pi i \rho(\theta_i)$ in \eqref{eq:large N SD} is introduced by identifying an even function $f(\theta)$ added to the effective action with the density $\rho(\theta)$, ensuring consistency with the droplet picture.
In contrast, we have obtained this term directly, stemming from the branch ambiguity that arises when rewriting the integral in \eqref{eq:SD equation finite N} as a contour integral in the complex $z$-plane.
This perspective offers a natural explanation for the droplet structure observed in \cite{Dutta:2016byx}.

One generally has to solve the saddle point equation under the assumption that the support of $\rho(\theta)$ lies in $[-\pi, \pi]$. 
Although the gap solution, namely, the solution with the support of $\rho(\theta)$ being a proper subset of $[-\pi, \pi]$ is nontrivial in general, 
the no-gap solution, namely, the solution with the support of $\rho(\theta)$ being the whole interval $[-\pi, \pi]$, is simply given by 
\begin{align}
  \rho(\theta) = 
    \frac{1}{2\pi} 
    + \frac{1}{\pi}
    \sum_{k=1}^\infty \beta_k \cos k\theta\,,
  \label{eq:no-gap solution}
\end{align}
from the Hilbert transformation
\begin{align}
  \dashint_{-\pi}^\pi d\theta' \cos k\theta \cot\frac{\theta-\theta'}{2}
  = 2\pi \sin k\theta\,,
  \quad 
  (k=1,2,\cdots)
\end{align}
and the condition $\int_{-\pi}^\pi d\theta \rho(\theta) = 1$.
In this case, 
one of the equations in \eqref{eq:large N SD imag} trivially becomes $x=-1$, while the nontrivial one becomes a linear equation for $x$ as 
\begin{align}
  x = 
  2\sum_{k=1}^\infty \beta_k \cos k\theta \,.
  \label{eq:linear boundary equation}
\end{align}

%%%
\subsection{Spectral curve and the density of the Maya diagram}

In \cite{Dutta:2016byx,Chattopadhyay:2017ckc}, it is discussed that the solution of the boundary equation \eqref{eq:boundary equation} with respect to $\theta$, which we denote by $\pm\theta(x)$, gives the density of the Maya diagram in the large $N_c$ limit as
\begin{align}
  \theta(x) = \pi \tilde{\rho}(x)\,.
  \label{eq:Maya density by theta}
\end{align}
We can understand this relation by regarding the boundary equation \eqref{eq:boundary equation} as a spectral curve of the model%
\footnote{
  It is interesting that essentially the same relation \eqref{eq:Maya density by theta} holds for the two-dimensional Yang-Mills theory on a cylinder \cite{gross1995some}. 
}. 

To this end, we rescale the variables $h_i$ to $\redx_i$ as \eqref{eq:maya diagram points} 
with $\epsilon = N_c^{-1}$ up to an irrelevant constant shift.
Since the variables $\redx_i$ and $\theta_i$ are related by the Fourier transformation in the ```path integral'' expression \eqref{eq:X_R by Fourier transform}, 
they are regarded as the eigenvalues of the canonical variables $\hat{\redx}_i$ and $\hat{\theta}_i$ which satisfy the canonical commutation relation of a quantum mechanics
with the Planck constant $N_c^{-1}$ as 
\begin{align}
  [\hat{\redx}_i, \hat{\theta}_j] = \frac{i}{N_c} \delta_{ij}\,.
\end{align}
Therefore, in the momentum representation, $\hat{\redx}_i$ can be expressed as the differential operator with respect to $\theta_i$ as
\begin{align}
  \hat{\redx}_i \equiv -\frac{i}{N_c}\frac{\del}{\del \theta_i}\,.
\end{align}

Since the functional $\tilde{\Xi}(\vec{\theta};\vec{\beta})$ satisfies the equations
\begin{align}
  \left(
    -\frac{1}{N_c}\frac{\del}{\del \theta_i}
    +
    i \sum_{k=1}^\infty \beta_k e^{ik\theta_i} 
    + \dashint_{-\pi}^\pi d\theta \rho(\theta)\cot\frac{\theta_i-\theta}{2}
    \pm 2\pi i \rho(\theta_i)
  \right) \tilde{\Xi}(\vec{\theta};\vec{\beta}) = 0\,,
  \label{eq:quantum curve pre}
\end{align}
after imposing the solution $\rho(\theta)$ of the saddle point equation \eqref{eq:large N SD real}, 
we can interpret \eqref{eq:quantum curve pre} as the Schr\"odinger-type equations
\begin{align}
  \hat{D}_\pm(\hat{\redx}_i, \hat{\theta}_i)\, \tilde{\Xi}(\vec{\theta};\vec{\beta}) = 0\,,
  \label{eq:quantum curve}
\end{align}
with 
\begin{align}
  \hat{D}_\pm (\hat{\redx}_i, \hat{\theta}_i) &\equiv
  - \hat{\redx}_i - \frac12
  + \sum_{k=1}^\infty \beta_k \cos k\hat\theta_i
  \pm 2\pi \rho(\htheta_i)\,,
\end{align}
where we have shifted $\hat{\redx}_i$ by the constant $\frac12$ to match the definition of $x$ in \eqref{eq:large N SD imag}.
The equations \eqref{eq:quantum curve} define a quantum curve of this system, 
and the equation \eqref{eq:large N SD} is regarded as the semi-classical limit of this quantum curve, that is, the spectral curve of the model realized in the large $N_c$ limit.

Since the variable $\redx_i$ represents the position of the $i$-th black circle in the Maya diagram,
this is a quantum-mechanical system of $N_c$ fermions, and the boundary equation \eqref{eq:boundary equation} defines the Fermi surface in the phase space spanned by $x$ and $\theta$.
Because one fermion occupies an area $2\pi \hbar = 2\pi N_c^{-1}$ in phase space, the droplet area mentioned in the previous subsection is understood as $2\pi N_c^{-1} \times N_c = 2\pi$.
From this viewpoint, we can evaluate the density of the Maya diagram at a fixed value of $x$.
Indeed, the number of fermions in a narrow strip of the droplet (Fermi sea) around $x$ with width $\delta x$ in phase space is given by $\frac{2\theta(x)\delta x}{2\pi N_c^{-1}}=N_c \frac{\theta(x)}{\pi} \delta x$.
This implies that the density of the Maya diagram at $x$ is given by \eqref{eq:Maya density by theta}.

Note that the reason why 
only $\hat{D}_+$ with the no-gap solution \eqref{eq:no-gap solution} is realized in \cite{Kimura:2020sud}
is that the Schur measure considered in \cite{Kimura:2020sud} does not have a restriction on the number of rows of the Young diagram by definition. 
In this case, the GWW phase transition does not occur
and it is natural that only $\hat{D}_+$ is realized as the quantum curve of the model.
In the case of the unitary matrix model with the action \eqref{eq:general model action},
the corresponding limiting shape of the Young diagram depends on the values of the coupling constants $\vec{\beta}$ and then the solution of the saddle point equation can be either the no-gap solution or the gap solution, which leads to the realization of both $\hat{D}_+$ and $\hat{D}_-$ as quantum curves of the model.

%%%
\subsection{Example: FKM model on the cycle graph}

To present a concrete example, let us check if the relation \eqref{eq:Maya density by theta} holds for the FKM model on the cycle graph $C_L$ where the coupling constants are given by $\beta_k = \gamma q^{kL}$.
In this case, by using the relation
\begin{align}
  \sum_{k=1}^\infty \beta_k e^{ik\theta} = \frac{\gamma q^L e^{i\theta}}{1-q^L e^{i\theta}} = \frac{\gamma \left(q^L \cos\theta - q^{2L} + i q^L \sin\theta\right)}{1-2q^L \cos\theta + q^{2L}} \,, 
\end{align}
the saddle point equation \eqref{eq:large N SD real} for the FKM model on $C_L$ becomes 
\begin{align}
    \dashint_{-\pi}^\pi d\theta' \rho(\theta')\cot\frac{\theta-\theta'}{2} 
    = 2\sum_{k=1}^\infty \gamma q^{kL} \sin k\theta 
    = \frac{2\gamma q^L \sin\theta}{1-2q^L \cos\theta + q^{2L}}\,, 
    \label{eq:saddle point equation for FKM}
\end{align}
and the boundary equation \eqref{eq:boundary equation} is given by
\begin{align}
  \left(x + \frac12 - \frac{\gamma \left(q^L \cos\theta - q^{2L}\right)}{1-2q^L \cos\theta + q^{2L}}\right)^2 = \pi^2 \rho^2(\theta)\,,
  \label{eq:boundary equation for FKM}
\end{align}
where $\rho(\theta)$ is the solution \eqref{eq:sln} of the saddle point equation \eqref{eq:saddle point equation for FKM}.

\begin{figure}[htbp]
  \begin{minipage}[b]{0.32\textwidth}
    \centering
    \includegraphics[width=\textwidth]{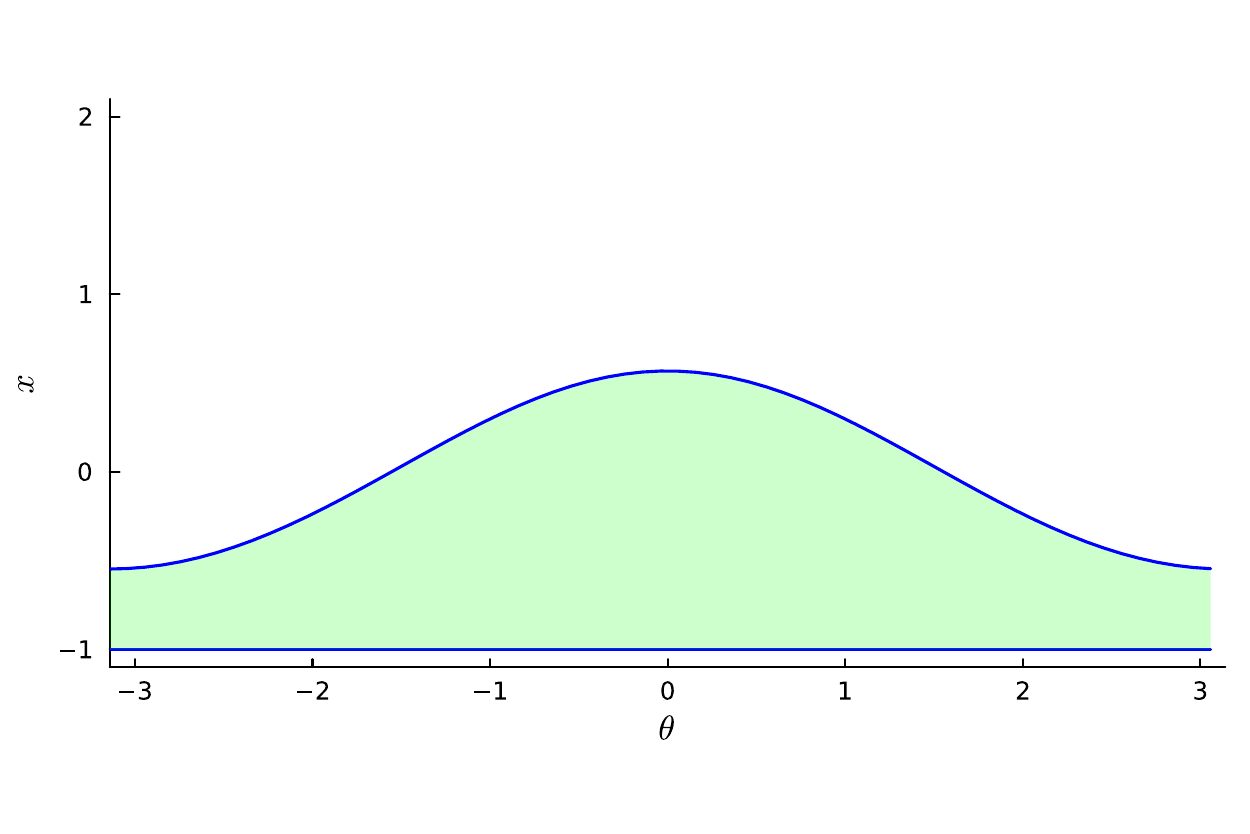}
    \subcaption{$q<q^*$}\label{subfig:droplet q<qc}
  \end{minipage}
  \begin{minipage}[b]{0.32\textwidth}
    \centering
    \includegraphics[width=\textwidth]{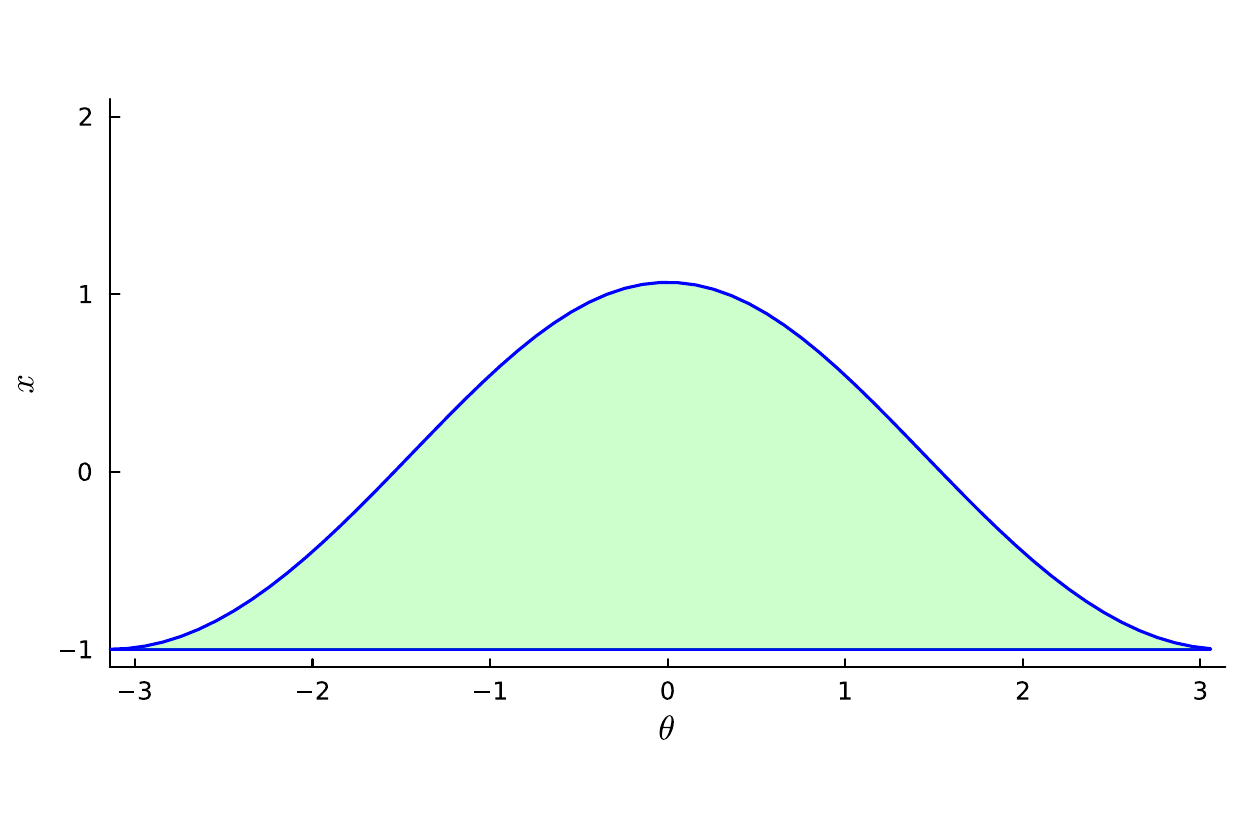}
    \subcaption{$q=q^*$}\label{subfig:droplet q=qc}
  \end{minipage}
  \begin{minipage}[b]{0.32\textwidth}
    \centering
    \includegraphics[width=\textwidth]{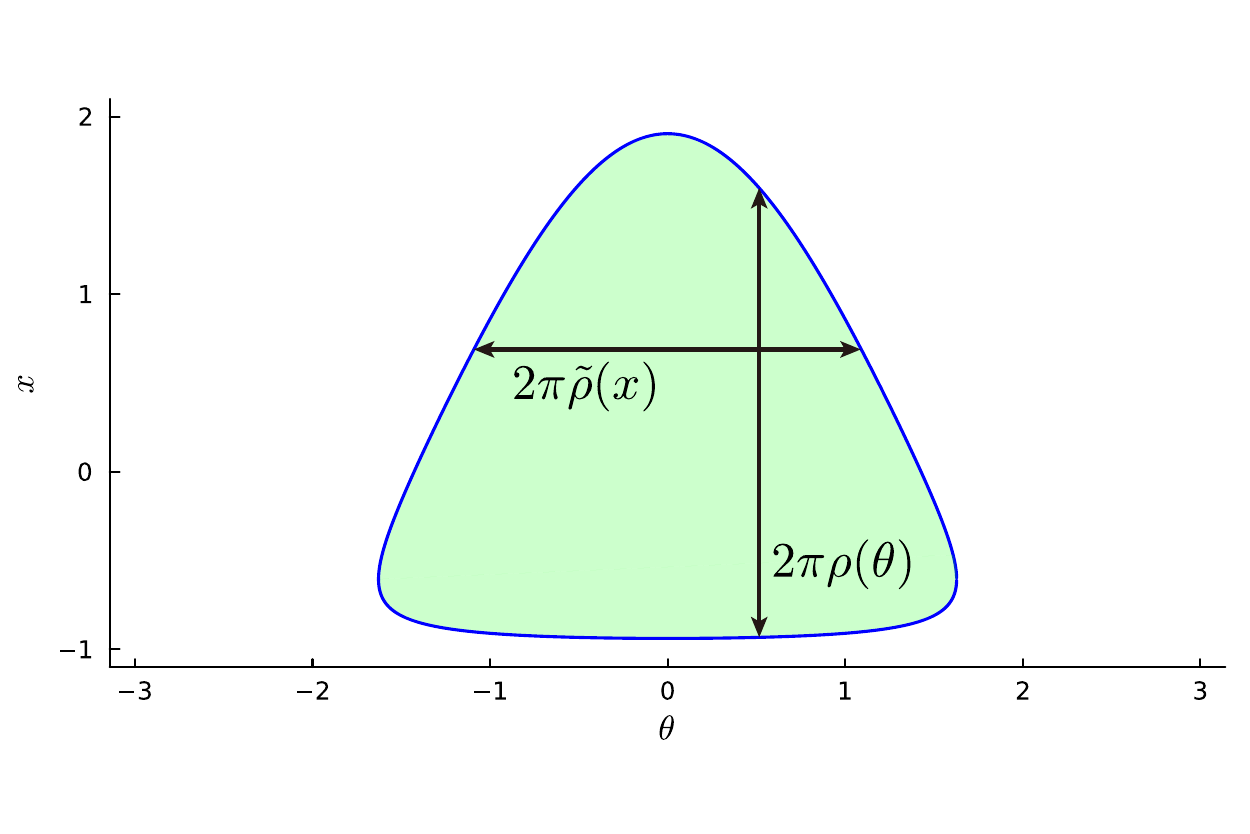}
    \subcaption{$q>q^*$}\label{subfig:droplet q>qc}
  \end{minipage}
\caption{The shape of the droplet defined by the boundary equation \eqref{eq:boundary equation for FKM} with $\gamma=16$ and $L=3$.}
\label{fig:droplets}
\end{figure}
The shape of the droplet \eqref{eq:boundary equation for FKM} is different depending on the value of $q$ and the transition between the no-gap phase and the gap phase occurs at $q=q^*$, 
which is shown in Fig.~\ref{fig:droplets} for various values of $q$ at fixed $\gamma$ and $L$.
When $q<q^*$, the droplet has a single connected component and the no-gap solution is realized. 
As mentioned in the previous subsection, the lower boundary of the droplet becomes $x=-1$ and only the upper boundary is non-trivial in this case. 
On the other hand, when $q>q^*$, the droplet is separated from the line $x=-1$ and the gap solution is realized. 

We can solve the boundary equation \eqref{eq:boundary equation for FKM} with respect to $\theta$ as $\theta=\pm \theta(x)$ with
\begin{align}
  \theta(x) =
  \begin{cases}
     \arccos\left(\frac{(q^L+q^{-L})x + 2\gamma q^L}{2(x+\gamma)}\right) 
   & (-\frac{2\gamma q^L}{1+q^{L}}<x<\frac{2\gamma q^L}{1-q^{L}})  \\
     \pi  & (-1 < x < -\frac{2\gamma q^L}{1+q^{L}} )
  \end{cases}
\end{align}
for $q<q^*$ and 
\begin{align}
 \theta(x) = 
  \arccos\left(\frac{(2x+2\gamma+1)^2q^L+(2x+1)^2q^{-L}-2(2\gamma-1)}{8(x+\gamma)(x+1)}\right)\,, 
\end{align}
for $q>q^*$. 
Comparing these expressions with the solutions \eqref{eq:rho small q} and \eqref{eq:rho large q} of the saddle point equation \eqref{eq:saddle point equation}, 
we can confirm that the relation \eqref{eq:Maya density by theta} is indeed satisfied in both phases.

\section{Conclusion and Discussion}

In this paper, we have established a comprehensive framework for KM-type gauge theories on graphs and explored their profound connections to the random partitions.
By introducing the concept of the graph bundles, we demonstrated that the partition functions of these gauge theories can be systematically described by using the Artin-Ihara $L$-functions, which provides a unified description for %various models on discrete structures.
the models constructed so far. 
By applying the Plancherel formula to the FKM model, namely the KM-type model with the scalar fields in the fundamental representation, we successfully reformulated the unitary matrix model as a random partition model governed by the Schur measure.
Our exact solutions in the large $N_c$ limit revealed that the GWW phase transition occurs precisely when the edge of the limiting shape of the Young diagram touches the boundary of the allowed representation space.
We clarified that this transition is related to BEC within the representation space. 
We also proved that the strong/weak coupling duality observed on the regular graphs has a natural combinatorial interpretation as the exchange between a Young diagram and its complement, reflecting the underlying functional equation of the Artin-Ihara $L$-function. 
Finally, we found a definitive relationship between the eigenvalue density of the unitary matrix model and the Maya diagram density of the random partitions. 
By deriving a droplet picture from the spectral curve, we showed that the correspondence between these two densities is fundamentally rooted in the phase space structure of the underlying free fermion system. 
These results provide a new perspective on the interplay between gauge theories, matrix models, and combinatorics.

The partition function of the KM-type model, originally written as a unitary matrix integral, reduces to a discrete sum over quantities labeled by Young diagrams. This kind of reduction is reminiscent of the Duistermaat-Heckman localization theorem \cite{Duistermaat:1982vw}. Indeed, some unitary matrix integrals associated with lattice gauge theory are known to be evaluated by localization techniques \cite{Matsuura:2014nga, Ohta:2021jze}, 
and the localization points are labeled by the Young diagrams in many cases \cite{Nekrasov:2002qd, Nekrasov:2003rj,Pestun:2009nn}.
From this perspective, it is natural to expect that the partition function of the KM-type model on a graph can be evaluated by the localization technique. 
The Artin-Ihara $L$-function on the graph is well established mathematically, and it seems to be compatible with supersymmetry, which is crucial in the localization theorem.
It is also interesting to investigate the KM-type model on a general graph, which is regarded as a multi-unitary matrix model, from the viewpoint of the localization technique.

%%%%%
As shown in Sec.~\ref{sec:duality}, 
the duality of the FKM model on the cycle graph can be understood as the exchange of a Young diagram $R$ with its complement $\tilde{R}$ within a rectangle of height $N_c$. 
This duality appears to reflect the ``stringy interpretation'' of the model, echoing the work of Gross and Taylor \cite{gross1993two} in the two-dimensional Yang-Mills theory, where such dualities describe the gauge theory in terms of branched covers and the topological exchange of flux configurations.
On the other hand, 
from the perspective of our model, it is somewhat curious that two-dimensional Yang-Mills theory exhibits a duality between a Young diagram and its complement. 
In fact, the FKM model reduces to the standard Wilson lattice gauge theory only in the regime where $\gamma \gg 1$ and the coupling $q$ is very small in this region, the dual transformation replacing $q$ with $q^{-1}$ breaks down. 
Furthermore, while the duality in the KM-type models is mathematically grounded in the functional equation of the $L$-function, 
no such functional equation holds for two-dimensional Yang-Mills theory. 
Nevertheless, it remains an intriguing problem to investigate why both theories exhibit the same kind of duality despite these fundamental differences.

In addition, the duality of the KM-type model is a universal property of the model on regular graphs which is rooted in the functional equation of the graph zeta function. 
Apart from the cycle graph, the KM-type model is a multi-unitary matrix model where the corresponding random partition picture is more comprehensive and the duality is expected to be realized in a more intricate manner.
It is also interesting to understand the duality of the KM-type model on a general regular graph from the viewpoint of the random partition picture.

%%%%%
The current solution of the FKM model on the cycle graph is governed by the Schur measure $|s_R(x)|^2$, which establishes a correspondence between the $U(N_c)$ character expansion and a free fermion droplet picture in phase space. 
It will be interesting to extend this framework to more general measures, such as the Jack and Macdonald measures, which correspond to $(q, t)$-deformations of the $U(N_c)$ character expansion.
For example, the GWW phase transition in the model with the Schur measure can be understood as a phenomenon where the droplet in phase space touches the boundary of the allowed region.
By considering the $(q, t)$-deformation, a novel phase structure is expected to merge which may lead deeper insights into the phase transitions of the dual unitary matrix model (gauge theory).

\section*{Acknowledgments}
The authors would like to thank
K.~Sakai, 
H.~Shimada and
H.~Watanabe
for useful discussions and comments.
This work is supported in part
by Grant-in-Aid for Scientific Research (KAKENHI) (C), Grant Number 23K03423 (K.~O.)
and 
by Grant-in-Aid for Scientific Research (KAKENHI) (C), Grant Number 26K07070 (S.~M.)

\appendix
\section{Exact Solution of the Density at Large $N_c$}
\label{app:density}

%%%
In this appendix, we solve the saddle point equation \eqref{eq:saddle point equation} at large $N_c$,
where the density $\rhot(x)$ is regarded as a continuous function of $x$.
The density $\rhot(x)$ has a finite support on the interval $a'< x < b$ with $b > 0$, where $\rhot(x)> 0$ and otherwise $\rhot(x)=0$.

\subsection{Solution for $0 < q \le q^*$}

When $q$ is smaller than $q^*$, we expect that the edge of the support $a'$ reaches $a'=-1$ and the density $\rhot(x)$ takes the value $1$ in the region $-1\le x \le a$ for some $a<0$. 
Therefore, the integral in the left-hand side of the saddle point equation \eqref{eq:saddle point equation} can be evaluated as
\begin{align}
  \begin{split}
  \dashint_{-1}^b dx'\, \frac{\rhot(x')}{x'-x}
  &= \dashint_{-1}^a dx'\, \frac{1}{x'-x}
  + \dashint_a^b dx'\, \frac{\rhot(x')}{x'-x} \\
  &= \log\left(\frac{x-a}{x+1}\right) 
  + \dashint_a^b dx'\, \frac{\rhot(x')}{x'-x} 
  \end{split}
\end{align}
and the saddle point equation reduces to
\begin{align}
 \dashint_{a}^b dx'\, \frac{\rhot(x')}{x'-x}
  = \log\left(\frac{x+\gamma}{x-a}\,q^L\right)\,,
  \label{eq:saddle point equation small q} 
\end{align}
with the boundary condition $\rhot(a)=1$, $\rhot(b)=0$ and 
\begin{align}
  \int_a^b dx \, \rhot(x) = -a\,.
\end{align}

In order to solve this equation, 
it is convenient to use the fact that the equation
\begin{align}
  \dashint_{a}^b dx'\, \frac{f(x')}{x'-x} = g(x)
\end{align}
can be solved as 
\begin{align}
  f(x) = \frac{\int_{a}^b dx'\, f(x')
  -  \dashint_{a}^b dx'\, \frac{\sqrt{(b-x')(x'-a)}}{x'-x} \frac{g(x')}{\pi}}{\pi\sqrt{(b-x)(x-a)}} 
  \label{eq:solution of the integral equation}
\end{align}
in general. 
Using the integral formulas
\begin{align}
  \frac{1}{\pi}\dashint_{-1}^1 dx' \frac{\sqrt{1-x'^2}}{x'-x} &= -x
  \label{eq:formula1}
\end{align}
and 
\begin{align}
  \begin{split}
  \frac{1}{\pi}\dashint_{-1}^1 dx' \frac{\sqrt{1-x'^2}}{x'-x} \log\left(1+\frac{x'}{c}\right)
  &= -c + \sqrt{c^2-1} - \left(1 + \log\frac{c+\sqrt{c^2-1}}{2c}\right)x \\
  &\quad + \sqrt{1-x^2}\arctan\left(\frac{cx+1}{\sqrt{c^2-1}\sqrt{1-x^2}}\right)\, ,
  \label{eq:formula2}
  \end{split}
\end{align}
for $c>1$ and $-1\le x \le 1$,
we obtain the solution of the saddle point equation \eqref{eq:saddle point equation small q} as
\begin{align}
  \rhot(x) = 
  \begin{cases}
    1 & (-1 \le x < a) \\
  %\frac{1}{2} - \frac{1}{\pi}\arctan\left(\frac{\alpha v(x)+1}{\sqrt{\alpha^2-1}\sqrt{1-v(x)^2}}\right)
  \frac{1}{\pi}\arccos\left(\frac{\alpha v(x)+1}{\alpha+v(x)}\right)
  = 
  \frac{1}{\pi}\arccos\left(\frac{(q^L+q^{-L})x + 2\gamma q^L}{2(x+\gamma)}\right)
  & (a \le x \le b)
  \end{cases} \, ,
  \label{eq:rho small q}
\end{align}
where we have defined 
\begin{align}
  \alpha \equiv \frac{q^L+q^{-L}}{2}
\end{align}
and 
\begin{align}
  v(x) = \frac{2}{b-a}\left(x-\frac{b+a}{2}\right)
  \label{eq:s by ab}
\end{align}
with 
\begin{align}
  a = \frac{-2\gamma q^L}{1+q^L}\,, \quad 
  b = \frac{2\gamma q^L}{1-q^L}\,.
  \label{eq:ab small q}
\end{align}
The condition $a\le -1$ is compatible with $q \le q^*$, 
which gives the range of $q$ where this phase is realized.
In other words, the phase transition point is given by $(q^*)^L = \frac{1}{2\gamma-1}$.

%%%
\subsection{Solution for $q^* < q < 1$}

We next consider the case where the value of $q$ is larger than $q^*$ and one of the edges of the density does not reach $x=-1$
and the density $\rhot(x)$ takes the positive value in the region $a \leq x \leq b$.
Assuming $\rhot(a)=\rhot(b)=0$,
we can again solve the saddle point equation \eqref{eq:saddle point equation} by using \eqref{eq:formula1} and \eqref{eq:formula2} as 
\begin{align}
  \rhot(x) &= 
  \frac{1}{\pi}\arctan\left(\frac{\beta_+ v(x)+1}{\sqrt{\beta_+^2-1}\sqrt{1-v(x)^2}}\right)
  -
  \frac{1}{\pi}\arctan\left(\frac{\beta_- v(x)+1}{\sqrt{\beta_-^2-1}\sqrt{1-v(x)^2}}\right) \nn \\
  &=
  \frac{1}{\pi}\arccos\left(
    \frac{(2x+2\gamma+1)^2q^L + (2x+1)^2q^{-L} - 2(2\gamma-1)}{8(x+1)(x+\gamma)}
  \right)\,,
  \label{eq:rho large q}
\end{align}
where the function $v(x)$ is again defined by \eqref{eq:s by ab} with
\begin{align}
  a = \frac{(2\gamma+1)q^L-1-2\sqrt{(2\gamma-1)q^L}}{2(1-q^L)}\,, \quad 
  b = \frac{(2\gamma+1)q^L-1+2\sqrt{(2\gamma-1)q^L}}{2(1-q^L)}\,, 
  \label{eq:ab large q}
\end{align}
and we have defined 
\begin{align}
  \beta_+ \equiv \frac{(2\gamma-1)q^L+1}{2\sqrt{(2\gamma-1)q^L}}\,, \quad
  \beta_- \equiv \frac{(2\gamma-1)q^{-L} + 1}{2\sqrt{(2\gamma-1)q^{-L}} }\,. 
  \label{eq:C1 Cgamma}
\end{align}
The condition $a\le -1$ means $q > q^*$,
which gives the range of $q$ where this phase is realized.
The phase transition point is again given by $(q^*)^L = \frac{1}{2\gamma-1}$.

\subsection{Internal energy}
\label{app:internal energy}

%namely the phase transition should occur at the point $q^* = \frac{1}{2\gamma-1}$. To see this phase transition from the random partition perspective, we need to evaluate the physical observables such as the internal energy and the specific heat.

Let us now evaluate the internal energy \eqref{eq:energy} of the model from the partition function \eqref{dual partition function}.

Since the derivative of $\dim_{N_f} R$ with respect to $\gamma$ is given by
\be
\frac{\del}{\del \gamma}\left(\dim_{N_f}R\right)
= \dim_{N_f}R \sum_{(i,j)\in R}\frac{1}{\gamma+\epsilon(j-i)}
\, ,
\ee
we obtain the derivative of the partition function (\ref{dual partition function}) with respect to $\gamma$ as follows:
\be
\frac{\del}{\del \gamma}Z_{\rm FKM}^{C_L}
= 2 \sum_{R\in Y_{N_c}}
\left(\dim_{N_f}R\right)^2
\left(
\sum_{(i,j)\in R}\frac{1}{\gamma+\epsilon(j-i)}
\right)
q^{2L|R|}\,. 
\ee
Then, we can evaluate the internal energy as
\be
  E_{C_L} = 
  -\frac{1}{N_c^2}\gamma \frac{\del }{\del \gamma}\log Z_{\rm FKM}^{C_L} 
  = -\frac{2\gamma}{N_c^2}\left\langle
    O_R
  \right\rangle \, ,
  \label{eq:internal energy from partition function}
\ee
%where $F_{C_L}$ is the free energy given by
%\be
%F_{C_L} = -\frac{1}{N_c^2}\log Z_{C_L}\, ,
%\ee and
where we have defined the operator $O_R$ as
\be
  O_R \equiv 
  %\sum_{(i,j)\in R}\frac{1}{\gamma+\epsilon(j-i)}
  \sum_{i=1}^{N_c}\sum_{j=1}^{\lambda_i} \frac{1}{\gamma + \epsilon(j-i)} \\
  \, .
  \label{operator OR}
\ee

%Let us next consider the expectation value of the operator $O_R$ in the large $N_c$ limit. 
Since $\epsilon=\frac{1}{N_c}$ is sufficiently small in the large $N_c$ limit, we can approximate the operator $O_R$ as 
\be
\begin{split}
O_R 
&\sim N_c\sum_{i=1}^{N_c} \log\frac{\gamma + \epsilon n_i}{\gamma - \epsilon i} \\
&\sim N_c^2 \left(
  \int_{-1}^\infty dx \, \rhot(x) \log\left(x+ \gamma\right)
\right)
-N_c^2 \left(\gamma\log\gamma-(\gamma-1)\log(\gamma-1)-1\right)
 \, ,
\end{split}
\ee
where we have used $n_i = \lambda_i -i+1$ and the integral approximation for small $\epsilon$.
In the large $N_c$ limit, the internal energy \eqref{eq:internal energy from partition function} can be evaluated by using the (continuous) density $\rhot(x)$ at the saddle point as 
\begin{align}
  \begin{split}
 E_{C_L} &= -2\gamma \int_{a'}^b dx\, \rhot(x)\log(x+\gamma)
 + 2\gamma\left(\gamma\log\gamma - (\gamma-1)\log(\gamma-1)-1\right) \\
 &= %2\gamma \int_{a'}^b dn\, \Bigl(\rho'(n) \left((n+\gamma)\log(n+\gamma)-n+a'\right)\Bigr) 
 I + \rhot(a')2\gamma(\gamma+a')\log(\gamma+a') 
 + 2\gamma^2\log\gamma - 2\gamma(\gamma-1)\log(\gamma-1)-2\gamma \, , 
 \label{eq:internal energy CL}
  \end{split}
\end{align}
where we have used a partial integral and $I$ is given by 
\begin{align}
  I \equiv 2\gamma \int_{a'}^b dx\, \left(\frac{d\rhot(x)}{dx} \left((x+\gamma)\log(x+\gamma)-x+a'\right)\right)\, ,
\end{align}
and $a'$ and $b$ are the endpoints of the support of the density $\rhot(x)$ depending on each phase.

%%%
\subsubsection*{Internal energy for $0 < q \le q^*$}

When $0 < q \le q^*$, the density $\rhot(x)$ is given by \eqref{eq:rho small q}.
Noting $\rhot'(x)=0$ for $-1\le x \le a$ in this case, 
the integral $I$ can be evaluated as
\begin{align}
  \begin{split}
  I &= -\frac{2\gamma^2}{\pi}\int_a^b \frac{dx}{\sqrt{(b-x)(x-a)}}
  \left(
    \log(x+\gamma) - \frac{x+1}{x+\gamma}
  \right) \\ 
  &= 
  2\gamma^2\log\left({1-q^{2L}}\right)
  - 2\gamma\left(\gamma\log\gamma -1\right)\,.
  \end{split}
\end{align}
Substituting this result  and $\rhot(a')=\rhot(-1)=1$ into \eqref{eq:internal energy CL}, we obtain the internal energy in this phase as 
\begin{align}
  E_{C_L} = 2\gamma^2 \log\left(1- q^{2L}\right)\,, 
  \label{eq:energy of random partition 1}
\end{align}
which reproduces \eqref{eq:energy Cn}.

%%%
\subsubsection*{Internal energy for $q > q^*$}

When $q > q^*$, the density $\rhot(x)$ is given by \eqref{eq:rho large q}.
In order to evaluate the integral $I$, it is convenient to define the function 
\begin{align}
  f_\beta(x) \equiv \frac{1}{\pi}\arctan\left(\frac{\beta v(x)+1}{\sqrt{\beta^2-1}\sqrt{1-v(x)^2}}\right)\,, 
\end{align}
where $v(x)$ is given by \eqref{eq:s by ab} with $a$ and $b$ given by \eqref{eq:ab large q}.
%with the parameter $s$ given by \eqref{eq:s by ab}. 
%Since the derivative of this function is given by
%\begin{align}
  %f_C'(s) = \frac{\sqrt{C^2-1}}{\pi}\frac{1}{(s+C)\sqrt{{1-s^2}}}\,, 
%\end{align}
Then we can separate the integral $I$ as $I=I_{\beta_1} - I_{\beta_\gamma}$ with 
\begin{align}
  I_\beta &\equiv 2\gamma \int_{a}^b dx\, \Bigl(\frac{df_\beta(x)}{dx} \left((x+\gamma)\log(x+\gamma)-x+a\right)\Bigr) \nn \\
  &= \gamma(b-a)\frac{\sqrt{\beta^2-1}}{\pi}\int_{-1}^1 \frac{dt}{\sqrt{1-t^2}}\frac{1}{t+\beta} 
  \left(
    (t+\beta_\gamma)\log\frac{(b-a)(t+\beta_\gamma)}{2} - (t+1)
  \right) \nn \\
  &= \gamma(b-a) \left(
    \left(\beta_\gamma-\beta+\sqrt{\beta^2-1}\right)\log\frac{b-a}{2} + \beta-\sqrt{\beta^2-1}-1
  \right) \nn \\
  &\quad + \gamma(b-a)\frac{\sqrt{\beta^2-1}}{\pi}\int_{-1}^1 \frac{dt}{\sqrt{1-t^2}}
  \frac{t+\beta_\gamma}{t+\beta}\log(t+\beta_\gamma)\,, 
\end{align}
where $\beta_\gamma = \frac{2\gamma-1+q^L}{2\sqrt{(2\gamma-1)q^L}}=\frac{b+a+2\gamma}{b-a}$. 
Using the integration formulas 
\begin{align}
  \begin{split}
  \frac{1}{\pi}\int_0^\pi d\theta \log\left(\cos \theta + c\right) 
  &= \log\left(\frac{c+\sqrt{c^2-1}}{2}\right)\,, \\ 
  \frac{1}{\pi}\int_0^\pi \frac{\log(\cos\theta+c_1)}{\cos\theta+c_2}
  &= \frac{1}{c_2^2-1}\log\left(
    \frac{c_1c_2-1 + \sqrt{(c_1^2-1)(c_2^2-1)}}{c_2+\sqrt{c_2^2-1}}
  \right)\,,
  \end{split}
\end{align}
for $|c|>1$, $|c_1|>1$ and $|c_2|>1$, we obtain 
\begin{align}
  \begin{split}
  I_{\beta_\gamma} = \gamma(b-a)\Biggl(
    &\sqrt{\beta_\gamma^2-1}\,\log\frac{b-a}{2} + \beta_\gamma-\sqrt{\beta_\gamma^2-1}-1 
    %\left(C_\gamma\log\frac{b-a}{2}-1\right) - \left(\log\frac{b-a}{2}-1\right)\left(C_\gamma-\sqrt{C_\gamma^2-1}\right)  \nn \\
   +\sqrt{\beta_\gamma^2-1}\,\log\frac{\beta_\gamma+\sqrt{\beta_\gamma^2-1}}{2}
  \Biggr)\,, \\
  I_{\beta_1}  = \gamma(b-a)\Biggl(
     &\left(\beta_\gamma-\beta_1+\sqrt{\beta_1^2-1}\right)\log\frac{b-a}{2} + \beta_1-\sqrt{\beta_1^2-1}-1  \nn \\
     & + \sqrt{\beta_1^2-1}\log\frac{\beta_\gamma+\sqrt{\beta_\gamma^2-1}}{2}\\
     & +(\beta_\gamma-\beta_1)\log\left(
      \frac{\beta_1\beta_\gamma-1 + \sqrt{(\beta_\gamma^2-1)(\beta_1^2-1)}}{\beta_1+\sqrt{\beta_1^2-1}}
     \right)
   \Biggr)
  \end{split}
\end{align}
Substituting these results into \eqref{eq:internal energy CL} with \eqref{eq:C1 Cgamma} and \eqref{eq:ab large q}, 
we obtain the internal energy in this phase as 
\begin{align}
  E_{C_L} &= I_{\beta_1} - I_{\beta_\gamma} + 2\gamma^2\log\gamma -2\gamma(\gamma-1)\log(\gamma-1) - 2\gamma\nn \\
  &= 2\gamma\log(1-q^L) + 2\gamma(\gamma-1)\log\left(1-\frac{1}{\gamma}\right) 
  -2\gamma(2\gamma-1)\log\left(1-\frac{1}{2\gamma}\right)\,, 
\end{align}
which again exactly agrees with the previous result \eqref{eq:energy Cn}.

\bibliographystyle{ptephy.bst}
\bibliography{refs}

\end{document}